\documentclass{article}

\usepackage[english]{babel}
\usepackage[T1]{fontenc}

\usepackage[letterpaper,top=2cm,bottom=2cm,left=3cm,right=3cm,marginparwidth=1.75cm]{geometry}

\usepackage{amsmath,amsfonts}
\usepackage{graphicx}
\usepackage[colorlinks=true, allcolors=blue]{hyperref}

\usepackage[normalem]{ulem}

\usepackage{xcolor}

\newcommand{\eps}{\epsilon}

\newcommand{\ph}{\phantom}
\newcommand{\nn}{\nonumber}

\usepackage{float} 
\usepackage{authblk}
\usepackage{orcidlink}
\usepackage{comment}

\title{A Cartan-geometrical perspective on torsion and non-metricity}

\author{
Damianos Iosifidis\orcidlink{0000-0003-1849-6481}$^{1}$\thanks{damianos.iosifidis@ut.ee}  \quad 
Tomi Koivisto\orcidlink{0000-0002-9403-8565}$^{2}$\thanks{tomi.koivisto@ut.ee}
\quad
Wit Kro\'{s}nicki$^{3}$\thanks{w.krosnicki.957@studms.ug.edu.pl}
\quad
Tom Zlosnik\orcidlink{0000-0001-7715-5842}$^{3}$\thanks{thomas.zlosnik@ug.edu.pl} 
}

\date{
\small
$^{1}$ Scuola Superiore Meridionale, Largo San Marcellino 10, 80138 Napoli, Italy and
INFN– Sezione di Napoli, Via Cintia, 80126 Napoli, Italy\\
$^{2}$ University of Tartu, \"Ulikooli 18, 50090 Tartu, Estonia\\
$^{3}$ Institute of Theoretical Physics and Astrophysics,
University of Gda\'{n}sk, Wita Stwosza 57, 80-308 Gda\'{n}sk, Poland
}

\begin{document}
\maketitle

\begin{abstract}
\'{E}lie Cartan established that the metric and intrinsic curvature of a $D$ dimensional embedded manifold $M$ could be determined by tracing the response of another surface $N$ of the same dimensionality, as it is rolled without slipping and twisting on $M$. In the context of spacetime geometry, this construction underpins the MacDowell-Mansouri formulation of General Relativity. We consider extensions of this framework that correspond to rolling of a shape with twisting and with shape evolution; it is shown that these naturally describe torsion and non-metricity respectively. Cartan-geometric formulations of teleparallel gravity and symmetric teleparallelism are discussed in addition to further manifestations of non-metricity in first-order formulations of gravity.
\end{abstract}

\section{Introduction}

The current prevailing theory of gravitation is Einstein's theory of General Relativity (GR). In GR, the gravitational field is identified with the metric tensor $g_{\mu\nu}$ of a four-dimensional pseudo-Riemannian manifold, with the familiar gravitational force being one of a wide variety of physical effects due to the non-vanishing of the Riemann curvature tensor $R^{\alpha}_{\ph{\alpha}\beta\mu\nu}$.

Despite the extraordinary success of GR, there are some reasons to reconsider the physical description of the gravitational field. On cosmological scales, it has been necessary to introduce dark energy and dark matter alongside $g_{\mu\nu}$ and the fields of the standard model of particle physics in order to account for observations. Though each of these is arguably relatively easily accommodated within GR (a cosmological constant can account for the evidence for dark energy whereas additional matter fields beyond the standard model provide a wide variety of dark matter candidates \cite{Albertus:2026fbe}), it is conceivable that dark energy and dark matter might reflect the need for an extension to GR. An example of this appears in the physics of the early universe, where data is consistent with there being a period of rapid expansion of space prior to the hot big bang. This can be ascribed to a new matter field or an effect of the non-minimal coupling of the standard model's Higgs field. However, a viable alternative description is a model due to Starobinsky \cite{Starobinsky:1980te} where an additional curvature term in the gravitational action results in $g_{\mu\nu}$ itself driving inflation.

There exist a number of gravitational models that describe the gravitational fields in terms other than just via the metric tensor. For example, if the quantity $\Gamma^{\alpha}_{\mu\nu}=\Gamma^{\alpha}_{\nu\mu}$ is regarded as an independent dynamical field, rather than a priori being the Levi-Civita connection $\bar{\Gamma}^{\alpha}_{\mu\nu}(g,\partial g)$, then from a specific Lagrangian - the Palatini Lagrangian - one can recover equations of motion for $\{g_{\mu\nu},\Gamma^{\alpha}_{\mu\nu}\}$ that are equivalent to Einstein's field equations of GR \footnote{Remarkably, there exist gravitational action principles which are functionals \emph{only} of $\Gamma^{\mu}_{\alpha\eta}$ and where the metric tensor only arises at the level of the equations of motion \cite{Bak:2022nrv}.}.

One can look to generalize this approach by allowing the field $\Gamma^{\alpha}_{\mu\nu}$ to be without any particular symmetry with respect to exchange of indices or via the anticipation that it may retain the same symmetry but be not entirely determined by the metric and its derivatives. The new possibilities created by a general $\Gamma^{\alpha}_{\mu\nu}$ can be encoded in the following tensors:

\begin{align}
T^{\alpha}_{\mu\nu} &= 2\Gamma^{\alpha}_{[\mu\nu]}\\
Q_{\mu\nu\alpha} &=  -\nabla^{(\Gamma)}_{\alpha}g_{\mu\nu}
\end{align}
which are respectively the torsion and non-metricity tensors. Both of which are zero if $\Gamma^{\alpha}_{\mu\nu} = \bar{\Gamma}^{\alpha}_{\mu\nu}(g,\partial g)$. Remarkably, there exist action principles where the Riemann curvature 

\begin{align}
R^{\alpha}_{\ph{\alpha}\beta\mu\nu} &=  2\partial_{[\mu|}\Gamma^{\alpha}_{\beta |\nu]} + 2\Gamma^{\alpha}_{\gamma [\mu|} \Gamma^{\gamma}_{\beta |\nu]}
\end{align}
is constrained to be zero, and yet the equations of motion correspond to the Einstein equations of GR, with effects usually attributed to spacetime curvature are either encoded entirely in torsion or entirely in non-metricity. The phenomenon by which gravitational effects can be attributed to a single one of $\{R^{\alpha}_{\ph{\alpha}\beta\mu\nu}, T^{\alpha}_{\mu\nu},Q_{\alpha\mu\nu}\}$ has been termed the \emph{Geometrical Trinity of Gravity}\footnote{General Parallel Relativity develops the geometrical trinity into a unified Hamiltonian theory of relativity, whose full quasi-local charge structure requires the underlying \(GL\) symmetry and uses torsion and non-metricity as boundary data for identifying physical charges such as entropy and angular momentum \cite{Koivisto:2026rmp}.} \cite{BeltranJimenez:2019esp}. 
The equivalence extends to the matter sector only when the matter action is formulated consistently with the corresponding affine geometry \cite{Iosifidis:2023eom}.


In all models within the geometrical trinity of gravity, gravitation is described by a metric tensor $g_{\mu\nu}$ alongside $\Gamma^{\alpha}_{\mu\nu}$. However, there exists an important set of models where $g_{\mu\nu}$ is itself in some sense a composite object. In the Einstein-Cartan formulation of gravity the `arena' of gravity is promoted from the spacetime manifold $M$ to being that of a vector bundle $E$ over $M$ with fibres $\mathbb{R}^{1,3}$ (i.e. equipped with a flat Minkowski metric $\eta_{IJ}=\text{diag}(-1,1,1,1)$). The formalism then introduces a new object: the co-tetrad/soldering form $e^{I}_{\mu}$ . This object is a spacetime one-form and transforms as a vector with respect to Lorentz transformations in the fibre at each point in spacetime. The metric $g_{\mu\nu}$ is then defined via the relation:

\begin{align}
g_{\mu\nu} &= \eta_{IJ}e^{I}_{\mu}e^{J}_{\nu} \label{geqee}
\end{align}
Indeed, the use of the vector bundle structure and field $e^{I}_{\mu}$ is necessary to couple fermions to the gravitational field. In addition, one can introduce a field $\omega^{I}_{\ph{I}J\mu}$ which is a spacetime one-form and is valued in the Lie algebra $\mathfrak{so}(1,3)$; analogously to the Palatini Lagrangian, there exists an action principle - polynomial in independent fields $e^{I}_{\mu}$ and $\omega^{IJ}_{\ph{IJ}\mu}$ which results in equations of motion equivalent to those of GR. This is the Einstein-Cartan formulation of gravity\footnote{Remarkably there exists an action principle for gravity which is a functional \emph{only} of $\omega^{I}_{\ph{I}J\mu}$ and not of $e^{I}_{\mu}$, with the metric recovered at the level of the equations of motion \cite{Krasnov:2011pp}. There are also interesting proposals where the metric emerges dynamically from spontaneous symmetry breaking of global $GL(n,R)$ down to its Lorentz subgroup \cite{Lindwasser:2022nfa}.}.

A natural generalization of the Einstein-Cartan model  is one in which the fibres of the vector bundle are no longer equipped with the Minkowski metric $\eta_{IJ}$.
Then the structure group is enlarged from  $SO(1,3)$  to that of the general linear group $GL(4,R)$  \cite{Percacci:2009ij}. The dynamical fields representing the gravitational field in this case can be taken to be a connection $A^{a}_{\ph{a}b\mu}$ valued in the Lie algebra $\mathfrak{gl}(4,R)$ and a soldering form $\theta^{a}_{\mu}$
- which transforms as a spacetime one-form and a $GL(4,R)$ vector - and a new \emph{dynamical} internal metric $\gamma_{ab}$, such that the spacetime metric tensor is to be identified with 

\begin{align}
g_{\mu\nu} &= \gamma_{ab}\theta^{a}_{\mu}\theta^{b}_{\nu}
\end{align}
In principle the matrix $\gamma_{ab}$ need not have Lorentzian signature but this is typically  imposed in order to recover a Lorentzian signature of $g_{\mu\nu}$.

It is furthermore possible to recover the connection $\Gamma^{\alpha}_{\mu\nu}$ - potentially containing non-vanishing torsion and non-metricity - as a solution of the following frame compatibility equation \cite{Hehl1995}:

\begin{align}
\partial_{\mu}\theta^{a}_{\nu} + A^{a}_{\ph{a}b\mu}\theta^{b}_{\nu} - \Gamma^{\alpha}_{\mu\nu}\theta^{a}_{\alpha}  &=0
\end{align}
In this sense, the fields $\{g_{\mu\nu},\Gamma^{\alpha}_{\mu\nu}\}$ can be recovered from $\{A^{a}_{\ph{a}b\mu},\gamma_{ab},\theta^{a}_{\alpha}\}$ and so one could look to recover the geometrical trinity of gravity as examples of a formulation of gravity based on $GL(4,R)$ gauge symmetry.

The action for gravity in Einstein-Cartan gravity has local $SO(1,3)$ symmetry. It was found \cite{MacDowell:1977jt,Stelle:1979aj} that one could alternatively formulate this theory as living on a vector bundle ${\cal E}$ over $M$ with fibres being either $\mathbb{R}^{1,4}$ or $\mathbb{R}^{2,3}$ (i.e. equipped with either the flat de Sitter metric $\eta^{(DS)}_{AB}=\mathrm{diag}(-1,1,1,1,1)$ or anti-de Sitter metric $\eta^{(ADS)}_{AB}=\mathrm{diag}(-1,1,1,1,-1)$). In such formulations, the dynamical variables for gravity are different: now the soldering form/co-tetrad itself is seen as a composite object, formed from a gauge field $A^{A}_{\ph{A}B\mu}$ in either the Lie algebra $\mathfrak{so}(1,4)$ or $\mathfrak{so}(2,3)$ and a compensator field $V^{A}$ such that $\eta_{AB}V^{A}V^{B}$ is assumed non-zero and fixed  (hence spontaneously breaking the $SO(1,4)$ or $SO(2,3)$ symmetry to $SO(1,3)$ if the norm is assumed positive in the former case and negative in the latter)  and the soldering form and metric are defined as:

\begin{align}
e^{A}_{\mu} &=  D_{\mu}V^{A} = \partial_{\mu}V^{A} + A^{A}_{\ph{A}B\mu}V^{B}\\
g_{\mu\nu} &= \eta_{AB}D_{\mu}V^{A}D_{\nu}V^{B}
\end{align}
In this sense, gravitation can be described as a spontaneously broken gauge theory based on groups $SO(1,4)$ or $SO(2,3)$.

The use of variables such as $\{A^{A}_{\ph{A}B\mu},V^{A}\}$ to describe the geometry of embedded surfaces was originally discovered by Cartan and 
corresponds to the determination of a geometry of a surface embedded in an ambient flat space by rolling another surface $N$ on it. For example, it can be shown that when rolling the two-sphere $S^{2}$ on a surface $M$ (here taken to be a two dimensional submanifold of $\mathbb{R}^{3}$) without twisting or slipping, then knowledge of the arc length inscribed on the sphere as it rolls along $M$ essentially allows the determination of the metric tensor on $M$, whilst the angle at which the sphere has rotated after rolling around a small loop is a measure of the difference between the Riemannian curvature of $M$ and the Riemannian curvature of $S^{2}$. The significance of $S^{2}$ is that it can be identified with the coset space $G/H$ - where here $G= SO(3)$ transformations describe the most general change of configuration of the sphere when rolled, whereas the $H=SO(2)$ subgroup of $SO(3)$ represent changes of configuration that preserve the point of contact between $M$ and $N$ (i.e. those transformations that do not correspond to rolling). Generalizations of this idea to the case where $M$ is a pseudo-Riemannian spacetime manifold have attracted interest for applications to gravitational theory \footnote{See for example \cite{Gryb:2012qt,Zlosnik:2018qvg,Koivisto:2025ryb,Gallagher:2022kvv,Addazi:2025vbw,Addazi:2026eto,Addazi:2026suu,Leclerc:2005qc}} as will be discussed in greater detail in Section \ref{section-macdowellmansouri} onward.

\begin{figure}
    \centering
    \includegraphics[width=0.6\linewidth]{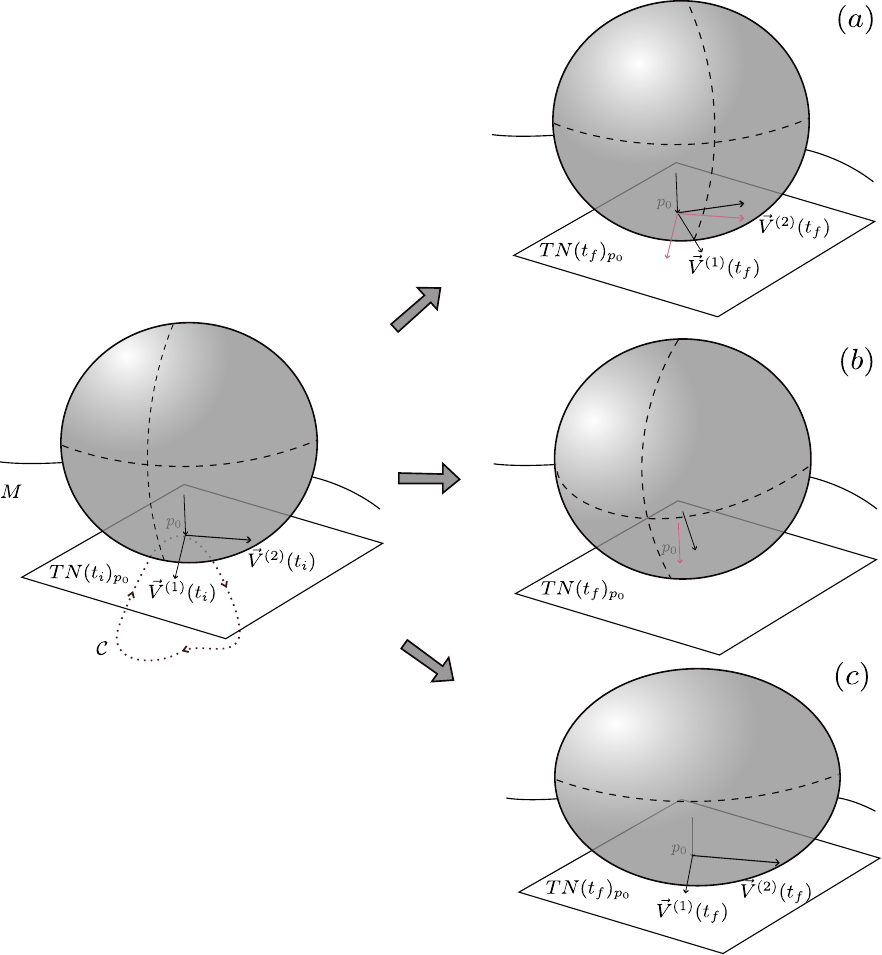}
    \caption{Three examples of transport of a shape $N$ on a manifold $M$ around a closed loop ${\cal C}$ beginning at a point $p_{0}$ on $M$ and with the curve parameterized by a parameter $t\in [t_{i},t_{f}]$ : a) The shape $N$ is a sphere which is rolled without slipping or twisting around ${\cal C}$. Upon returning to the starting point, there will generally have been a rotation of the sphere in a plane perpendicular to a vector in the tangent plane at $p_{0}$ with the amount of rotation determined by the difference in intrinsic curvatures between $M$ and $N$; b) the shape $N$ is a sphere which is rolled without slipping but with twisting, resulting rather in a `tilt' of the sphere so that the vector normal to the surface at $p_{0}$ is rotated about an axis lying in the tangent plane $TN(t_{i})_{p_{0}}$. This implies that a closed path on $M$ will not correspond to a closed path subtended on $N$. The change in the normal vector may be shown to encode information about the geometry of $M$ and this may be interpreted as a teleparallel connection; c) the shape $N$ at $t_{i}$ is a sphere but may evolve shape into an ellipsoid when transported without slipping around ${\cal C}$. Upon returning again to $p_{0}$ at $t=t_{f}$, the shape differs from its form at $t_{i}$ and this change is to encode parts of the geometry of $M$. This may be interpreted as being a consequence of non-metricity of the connection.}
\label{fig:rollingpossibilities}
\end{figure}

In this paper we will develop a Cartan-geometrical description of the different models within the generalized Metric-Affine Geometry. It will be shown that torsion may be interpreted as a specific twisting of $N$ as it is rolled without slipping on $M$ whereas non-metricity may be interpreted as a specific \emph{evolution of the shape} of $N$ as it is rolled on $M$. An illustration of these various possibilities is given in Figure \ref{fig:rollingpossibilities}. In turn, this evolution of the shape of $N$ may be interpretable as a change of section of a higher dimensional shape ${\cal N}$ transported on $M$; this higher dimensional shape is the coset space $G/H$ where $G$ represents that group of transformations of configuration of the shape when rolled/transported and $H$ represents the subgroup of $G$ that preserves the point of contact between $N$ and $M$.
This provides a Cartan-geometrical realization of the well-known interpretation of non-metricity as the non-preservation of lengths and inner products under parallel transport \cite{Hehl1995}.  Whether the present variables may also prove useful in describing possible matter couplings to non-metricity, through the putative dilation and shear components of matter's hypermomentum \cite{Hehl1976a,Hehl1976b}, remains an interesting question. For example, for the rolling of a two-dimensional shape \(N\) on a two-dimensional manifold \(M\), shape evolution is incorporated by enlarging the rolling group from \(SO(3)\) to either \(SL(3,R)\), if the volume of \(N\) is preserved, or \(GL(3,R)\), if it is not.


The structure of the paper is as follows: In Section \ref{section:ellipse} a simple lower dimensional example is considered: that of rolling a closed one dimensional manifold without slipping on a line. Here we introduce a generalization of rolling a fixed shape without slipping, establishing that shear of the shape can also contribute to what may more generally be termed \emph{transport without slipping}. In Section \ref{generalizedtransport} the mathematical description of transport without slipping of a higher dimensional shape $N$ on a shape $M$ of equal number of dimensions is established. 

In Section \ref{section-macdowellmansouri} we review the application of Cartan-gemetrical ideas to gravity via MacDowell-Mansouri gravity, wherein the metric and Riemannian curvature of a spacetime manifold $M$ can be described in terms of the rolling without slipping or twisting of a fixed spacetime $N$, with both locally embedded in a higher dimensional flat spacetime. In Section \ref{torsiontele}, this perspective is re-examined but with the condition of rolling without twisting being relaxed; specifically, it is shown how a specific recipe for twisting corresponds to torsion of the rolling connection as it appears in the teleparallel formulation of General Relativity. 

In Section \ref{section-nonmetricity}, the generalization of rolling without slipping to transport without slipping including shape evolution of $N$ is considered and it is shown that this is naturally interpreted as non-metricity of the rolling connection. Within this section, three different models are considered, each of which, as a classical theory of gravity, is dynamically equivalent to General Relativity and each may be interpreted from a Cartan-gemetrical viewpoint as involving shape evolution: the first (Section \ref{subsection-symtele}) corresponds to symmetric teleparallelism, wherein the rolling connection possesses no curvature but possesses non-metricity; the second (Section \ref{subsection-poincare}) is closely related to models where gravity is interpreted as a gauge theory of the Poincar\'{e} group; and the third (Section \ref{subsection-geometrodynamics}) is a model of geometrodynamics where - in a $3+1$ decomposition of spacetime - the change in the spatial metric when parallel transported around a closed curve on a surface of constant time is related to the trace-free part of that surface's extrinsic curvature in spacetime.
Finally, in Section \ref{section-discussionandconclusions} we present our conclusions and the scope for future work.

\section{Example: transport of an ellipse on a line}
\label{section:ellipse}

For illustrative purposes, we first consider a lower dimensional example: that of a one dimensional manifold $N$ being rolled along the manifold $M$ taken to be a line $L$ in two dimensional flat Euclidean space. Given a set of Cartesian coordinates $(x,y)$, for simplicity we take $L$ to be the line $y=-R$, for constant $R$.


We restrict the transport of $N$ so that at the point of contact between $N$ and $L$, their tangent spaces are aligned (or, equivalently here, the normal vectors to the surfaces are parallel and point in opposite directions). We also will constrain the transport process to be one of `transport without slipping' meaning that in moving from a point $p_{1}$ on $L$ to a nearby point $p_{1}+\delta p$, an arc length of $\delta p$ (to leading order in smallness) is subtended on the transported (evolving) shape. We parameterize the path on $L$ by a parameter $t$, which is taken to measure proper length along the line. Then, we have the following `evolution equation' for points $X^{A}(t)$ on $N$:

\begin{align}
\tilde{X}^{A}(t_{1}) &=  \Omega^{A}_{\ph{A}B}(t_{1},t_{0})X^{B}(t_{0}) + P^{A}(t_{1},t_{0}) \label{xtran}
\end{align}
where $\tilde{X}^{A}(t_{1})$ represents an embedding coordinate on $N$ and $\Omega^{A}_{\ph{A}B}\subset SL(2,R)$ with $\Omega^{A}_{\ph{A}B}(t_{0},t_{0})=\delta^{A}_{\ph{A}B}$ and $P^{A}(t_{0},t_{0})=0$. We will assume $\{\Omega^{A}_{\ph{A}B},P^{A}\}$ are independent of $X^{A}(t)$ so that generally both $X^{A}(t_{0})$ and $\tilde{X}^{A}(t_{1})$ will describe ellipses. What does this mean? This means that in its `rest frame', the ellipse $N$ undergoes an $SL(2,R)$ transformation of its points - and as such, consists of a combination of rotations and shears. The quantity $P^{A}$ describes how the `centre of mass' of $N$ moves in $\mathbb{R}^{2}$ as it is transported. Note that this may involve both horizontal and vertical changes of the position of the centre of mass.

The goal then is to understand what forms  $\{\Omega^{A}_{\ph{A}B},P^{A}\}$ take in different `transport without slipping' scenarios. For example, it will be shown that two options include transport purely by rolling (rotation) and transport produced entirely by shear and without rotation. For illustrative purposes, let's begin the process at $t_{0}$  with the shape being described as a circle:

\begin{align}
\delta_{AB}X^{A}X^{B} = R^{2} \label{sdef}
\end{align}
where $\delta_{AB}$ is the flat two dimensional Euclidean metric. Using  (\ref{xtran}) alongside (\ref{sdef})  we can then determine the shape coordinates $\tilde{X}^{A}$ on $N$ at $t_{1}$ as:

\begin{align}
\xi_{AB}(\tilde{X}^{A}-P^{A})(\tilde{X}^{B}-P^{B}) &= R^{2} \label{yxxeq}
\end{align}
where

\begin{align}
\xi_{AB} &=  \delta_{CD} (\Omega^{-1})^{C}_{\ph{C}A}(\Omega^{-1})^{D}_{\ph{D}B}
\end{align}
Firstly, we want to ensure that the minimum value of $\tilde{X}^{y}$ is $-R$. To this end, we can write (\ref{yxxeq}) in the form $F(X^{A}) = R^{2}$. Then, $\tilde{\nabla}_{A}F = \frac{\partial F}{\partial \tilde{X}^{A}}=2\xi_{AB}(\tilde{X}^{B}-P^{B})$ will tell us the direction of the normal vector at each point. 
We would like to have that at the point of contact $\tilde{X}^{y}_{(c)}=-R$ (where the subscript means `at the point of contact' for some value of $t$) we have $\tilde{\nabla}_{x}F=0$ so that $N$ is at its lowest point, so:

\begin{align}
0 &= 2\xi_{xx}(\tilde{X}^{x}_{(c)}-P^{x}) + 2\xi_{xy}(\tilde{X}^{y}-P^{y})
\end{align}
Now require that $\tilde{X}^{y}_{(c)}= -R$ which implies that at the point of contact

\begin{align}
\tilde{X}^{x}_{(c)} &=  \frac{\xi_{xy}}{\xi_{xx}}(R+P^{y}) + P^{x} \label{xwhere}
\end{align}
Using this in equation (\ref{yxxeq}), the equation for the transported surface, we obtain:

\begin{align}
\frac{(P^{y}+R)^{2}\det(\xi)}{\xi_{xx}} &= R^{2}
\end{align}
Given that $\det{\Omega}=1$, we have $\det{(\Omega^{-1})}=1$ and $\det{\xi}=1$ so 

\begin{align}
    P^{y} &= -R(1-\sqrt{\xi_{xx}})
\end{align}
Putting this solution into (\ref{xwhere}) we have:

\begin{align}
\tilde{X}^{x}_{(c)} &= P^{x}+ \frac{\xi_{xy}}{\sqrt{\xi_{xx}}}R
\end{align}
Next we require that at some time $t_{1}$ we have $\tilde{X}^{y}_{(c)}= - R$, $\tilde{X}^{x}_{(c)}= t_{1}-t_{0}$ so we find:

\begin{align}
P^{x} &= (t_{1}-t_{0})  - \frac{\xi_{xy}}{\sqrt{\xi_{xx}}}R
\end{align}
Therefore, we have ensured that $N$ is transported along the line $y=-R$ so that the point of contact is on the line and moves with `unit' speed in the positive $x$ direction.

The next requirement is that the transport is `without slipping'. In this case this is interpreted as - for each moment $t$ along the path -  taking the `velocity' of the change of contact point to have `unit norm' in terms of rate of change of position of the contact point on $L$. That is to say, the rate at which distance is rolled along $N$ should - at each moment - match the rate of distance rolled along $L$.
As an application to one dimensional manifolds of the more general result (\ref{ghnA}) applicable to the case of $d$ dimensions obtained in Section \ref{generalizedtransport}, the condition reduces to

\begin{align}
1 &=  \frac{1}{\kappa_{p}}\frac{d J_{yx}}{dt} \label{noslip}
\end{align}
where $J^{A}_{\ph{A}B} = (\Omega)^{C}_{\ph{C}B} d((\Omega^{-1})^{A}_{\ph{A}C})/dt$ and $\kappa_{p}=\xi_{xx}^{3/2}/R$ is the extrinsic curvature of $N$ at the point of contact between $N$ and $L$. We now consider several different possibilities for transport without slipping.

\subsection{Rotation transport}
\label{rot_tran}
\begin{figure}[h!]
    \centering
    \includegraphics[width=0.5\linewidth]{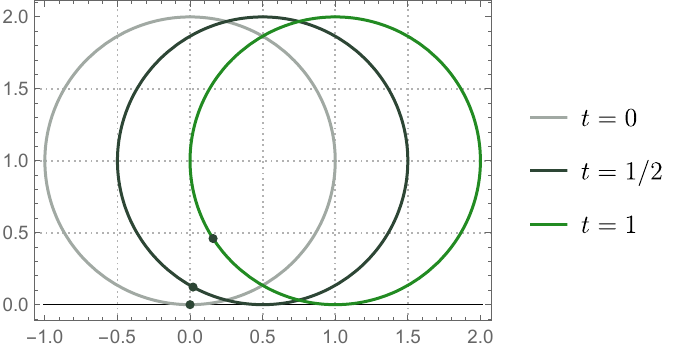}
    \caption{Rolling of a shape $N$ without slipping on a line $L$ (for ease of illustration taken to be $y=0$) under pure rotation. The black point denotes the location of the point on $N$ which was originally at $(x=0,y=0)$ at $t=0$, and then at subsequent times.}
        \label{fig:1}
\end{figure}

In this case, the shape $N$ is transported along $L$ by rotation. Consider some moment $t_{p}$. If the shape $N$ has configuration determined by $\Omega^{A}_{\ph{A}B}(t_{p})$ then we have

\begin{align}
\Omega^{A}_{\ph{A}B}(t_{p}+\delta t) &=  (\delta^{A}_{\ph{A}C}+ \zeta^{A}_{\ph{A}C}\delta t)\Omega^{C}_{\ph{A}B}(t_{p}) 
\end{align}
where

\begin{align}
\zeta^{A}_{\ph{A}B}(t) &= \begin{pmatrix}
0 & r(t) \\
-r(t) & 0
\end{pmatrix}
\end{align}
Which, given the initial condition $\Omega^{A}_{\ph{A}B}(t_{0}) = \delta^{A}_{\ph{A}B}$, has the solution:

\begin{align}
\Omega^{A}_{\ph{A}B}(t) &= \begin{pmatrix}
\cos(\phi(t)) & \sin(\phi(t)) \\
-\sin(\phi(t)) & \cos(\phi(t))
\end{pmatrix}
\end{align}
where $\phi(t)=\int^{t}_{t_{0}} r(t) dt$. Additionally $\xi_{AB} = \delta_{AB}$ due to the orthogonality of $\Omega^{A}_{\ph{A}B}$ in this case, then from equation (\ref{noslip}) we recover:

\begin{align}
\phi(t) &= \frac{t-t_{0}}{R}
\end{align}
This transport is shown in Figure \ref{fig:1}.

This is essentially the one dimensional application of Cartan's approach to geometry of embedded manifolds in terms of how a fixed, highly symmetric space responds when  rolled without slipping on them. In this case, the no-slipping condition guarantees that the distance subtended on $N$ as it rolls on $M$ is equal to how much distance is indeed traversed on $M$. 

\subsection{Shear transport}
\label{sheartransport}
\begin{figure}[h!]
        \centering
    \includegraphics[width=0.5\linewidth]{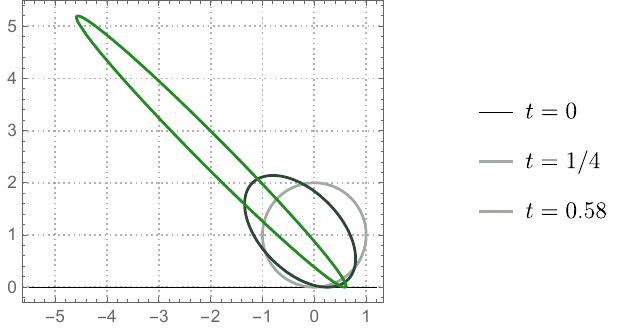}
        \caption{Illustration of shear transport without slipping on a line. The shape $N$ is originally a circle at $t=0$ and evolves to an ellipse of ellipticity that increases with $t$, and in so doing it can advance along the line $y=0$.}
        \label{fig:2}
\end{figure}

We now consider an alternative possibility, wherein the shape advances along $L$ by a shear-driven evolution of its shape and entirely without rolling. If the shape $N$ has configuration determined by $\Omega^{A}_{\ph{A}B}(t_{p})$ then we have

\begin{align}
\Omega^{A}_{\ph{A}B}(t_{p}+\delta t) &=  (\delta^{A}_{\ph{A}C}+ \zeta^{A}_{\ph{A}C}\delta t)\Omega^{C}_{\ph{A}B}(t_{p}) 
\end{align}
where

\begin{align}
\zeta^{A}_{\ph{A}B}(t) &= \begin{pmatrix}
0 & s(t) \\
s(t) & 0
\end{pmatrix}
\end{align}
Given the initial condition $\Omega^{A}_{\ph{A}B}(t_{0}) = \delta^{A}_{\ph{A}B}$, the solution is

\begin{align}
\Omega^{A}_{\ph{A}B}(t) &= \begin{pmatrix}
\cosh(\alpha(t)) & \sinh(\alpha(t)) \\
\sinh(\alpha(t)) & \cosh(\alpha(t))
\end{pmatrix}
\end{align}
where $\alpha(t) = \int^{t}_{t_{0}}s(t)dt$. Now from (\ref{noslip}) we have that:

\begin{align}
1 &= -\frac{R}{\cosh(2\alpha)^{\frac{3}{2}}}\frac{d\alpha}{dt}\label{sheartransp}
\end{align}
The solution with boundary condition $\alpha(t_{0})=0$ is plotted in Figure \ref{fig:2}. The shape advances by shearing so that the increasingly high ellipticity $N$ advances in the positive $x$ direction.

We see then that in this case, the evolving shape $N(t)$ may be considered to `transport' without slipping along $M$ by evolving its shape. Later, it will be shown to be an instance of \emph{non-metricity} of the rolling connection. Note that (\ref{sheartransp}) implies that an infinite range of $\alpha$ corresponds to a finite change in $t$ ($\sim 0.59 R$), implying that the amount of $L$ that can be traversed by this method is finite.

It is also possible to transport without slipping by a combination of shape-evolution/non-metricity and rolling, as we will now see.

\subsection{Combined rotation and shear transport}
\label{combinedrotationshear}
\begin{figure}[h!]
    \centering    \includegraphics[width=0.35\linewidth]{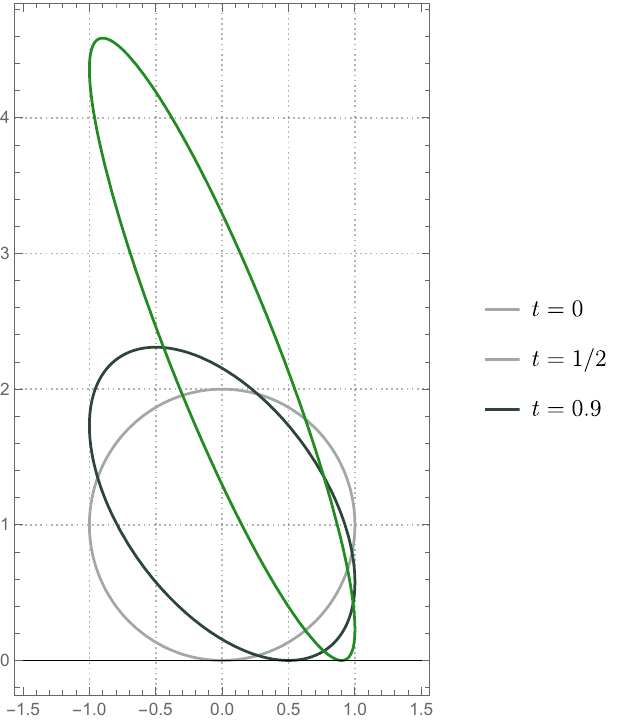}
    \caption{Combined shear and rotation transport without slipping on a line.}
    \label{fig:3}
\end{figure}

A special case is the following:

\begin{align}
\Omega^{A}_{\ph{A}B}(t_{p}+\delta t) &=  (\delta^{A}_{\ph{A}C}+ \zeta^{A}_{\ph{A}C}\delta t)\Omega^{C}_{\ph{A}B}(t_{p}) 
\end{align}
where

\begin{align}
\zeta^{A}_{\ph{A}B}(t) &= \begin{pmatrix}
0 & 0 \\
b(t) & 0
\end{pmatrix}  = \begin{pmatrix}
0 & \frac{b(t)}{2} \\
\frac{b(t)}{2} & 0
\end{pmatrix} +  \begin{pmatrix}
0 & \frac{-b(t)}{2} \\
\frac{b(t)}{2} & 0
\end{pmatrix}  \label{zetadef}
\end{align}
Hence, this can be interpreted as a combined infinitesimal shear and rotation. Given the initial condition $\Omega^{A}_{\ph{A}B}(t_{0}) = \delta^{A}_{\ph{A}B}$, there is the solution

\begin{align}
\Omega^{A}_{\ph{A}B}(t) &= \begin{pmatrix}
1  &  0 \\
B(t) & 1
\end{pmatrix}
\end{align}
where $B(t) = \int^{t}_{t_{0}}b(t) dt$. The no-slipping condition then takes the form:

\begin{align}
1 &=  \frac{R}{(1+B^{2})^{\frac{3}{2}}}\frac{dB}{dt}
\end{align}
with initial condition $B(t_{0})=0$. This equation has solution:

\begin{align}
B(t) &=  \frac{t-t_{0}}{\sqrt{R^{2}-(t-t_{0})^{2}}}
\end{align}
illustrating that this transport is capable of covering a path on $L$ of length equal to $R$, the radius of the original circle shape.
The resulting transport is plotted in Figure \ref{fig:3}.


The transpose of the matrix (\ref{zetadef}) is in the Lie algebra of the inhomogeneous orthogonal group $ISO(1) \subset SL(2,R)$ and we will see in Section \ref{subsection-poincare} that this simple lower dimensional example extends to a higher dimensional example of a shape transported via a combined shear and rolling such that the `rolling connection' - i.e. the higher dimensional generalization of the transpose of $(\ref{zetadef})$ - is valued in the four-dimensional Poincar\'{e} group $ISO(1,3) \subset SL(5,R)$, allowing for an interpretation of gravity as having a local Poincar\'{e} gauge symmetry.

\subsection{Interpretation in terms of transport of a higher dimensional shape}
The transport of an evolving ellipse can instead be described in terms of the transport of a two dimensional surface on the line, with the conic section at $t_{1}$ defining the ellipse (\ref{yxxeq}) as follows: 

Given Cartesian coordinates in $\mathbb{R}^{3}$ $(x,y,z)$ we take the line the ellipse is rolling on to be located at $y=-R,z=0$. We denote points on the cone as an embedded surface in $\mathbb{R}^{3}$ to be ${\cal Y}^{{\cal I}}$. It is is assumed to be a circular conic section $\delta_{AB}X^{A}X^{B}=R^{2}$ and that this circle can be regarded as points on the intersection with the plane $z=0$ where the cone is defined by:

\begin{align}
\eta_{{\cal I}{\cal J}}({\cal Y}^{{\cal I}}-{\cal Y}_{(0)}^{{\cal I}})({\cal Y}^{{\cal J}}-{\cal Y}_{(0)}^{{\cal J}})=0
\end{align}
where at $t_{0}$ $\eta_{{\cal I}{\cal J}}=\mathrm{diag}(1,1,-1)$ and ${\cal Y}_{(0)}^{{\cal I}}=(0,0,R)$. We now transport the cone along the line by a combined $SO(3)$ rigid body rotation and translation so that at a later time $t_{1}$ it has embedding coordinates $\tilde{\cal Y}^{{\cal I}}(t_{1})$ where

\begin{align}
\tilde{\cal Y}^{{\cal I}}(t_{1}) &= \Lambda^{{\cal I}}_{\ph{{\cal I}}{\cal J}}(t_{1},t_{0})({\cal Y}^{{\cal J}}(t_{0})-{\cal Y}^{{\cal J}}_{(0)}) + {\cal P}^{{\cal I}}(t_{1},t_{0})
\end{align}
for $\Lambda^{{\cal I}}_{\ph{{\cal I}}{\cal J}}\subset SO(3)$ and $\Lambda^{{\cal I}}_{\ph{{\cal I}}{\cal J}}(t_{0},t_{0})=\delta^{{\cal I}}_{\ph{{\cal I}}\cal J}$ and ${\cal P}^{{\cal I}}(t_{0},t_{0}) = {\cal Y}^{{\cal I}}_{(0)}$. Therefore at $t_{1}$ the surface of the cone satisfies:


\begin{align}
w_{{\cal I}{\cal J}}(\tilde{\cal Y}^{{\cal I}}-{\cal P}^{{\cal I}})(\tilde{\cal Y}^{{\cal J}}-{\cal P}^{{\cal J}})=0 \label{calYeq}
\end{align}
where

\begin{align}
w_{{\cal I}{\cal J}} &=  \eta_{{\cal K}{\cal L}}(\Lambda^{-1})^{{\cal K}}_{\ph{{\cal K}}{\cal I}}(\Lambda^{-1})^{{\cal L}}_{\ph{{\cal L}}{\cal J}}
\end{align}
Then restricting to the plane $z=0$ we have $\tilde{\cal Y}^{{\cal I}}=(\tilde{X}^{A},0)$ and we can write (\ref{calYeq}) as:


\begin{align}
\frac{1}{C}w_{AB}(\tilde{X}^{A}-V^{A})(\tilde{X}^{B}-V^{B}) &= R^{2}
\end{align}
where

\begin{align}
V^{A} &=  {\cal P}^{A} + (w^{-1})^{AI}w_{Iz}{\cal P}^{z}\\
C &= -\frac{1}{R^{2}}\bigg(w_{zz}+(w^{-1})^{IJ}w_{Iz}w_{Jz}\bigg){\cal P}^{z}{\cal P}^{z}
\end{align}
Now compare this to the equation describing the ellipse at $t_{1}$ (\ref{yxxeq}) we have that:

\begin{align}
\xi_{AB} &= \frac{1}{C}w_{AB}\\
P^{A} &=  V^{A}
\end{align}
The original picture of Rotation Transport presented in Section \ref{rot_tran} is an example of the original construction due to Cartan wherein the geometry of a Riemannian $m$-dimensional manifold was probed by rolling on it the \emph{fixed} model space $N$ realized as an embedding of $SO(m+1)/SO(m) \cong S^{m}$ into $\mathbb{R}^{m+1}$ as a sphere of a certain radius; the shape is transported along $M$ by an $SO(m+1)$ transformation, and $SO(m)$ can be interpreted as the stabilizer $\mathrm{Stab}(\vec{n})$ of a non-vanishing vector $n^{A}$ normal to the shape $N$ at the point of contact with $M$. By the orbit-stabilizer theorem, then $S^{m}$ is identified with the orbit of $n^{A}$ under $SO(m+1)$.

\begin{figure}[h!]
    \centering
\includegraphics[width=0.5\linewidth]{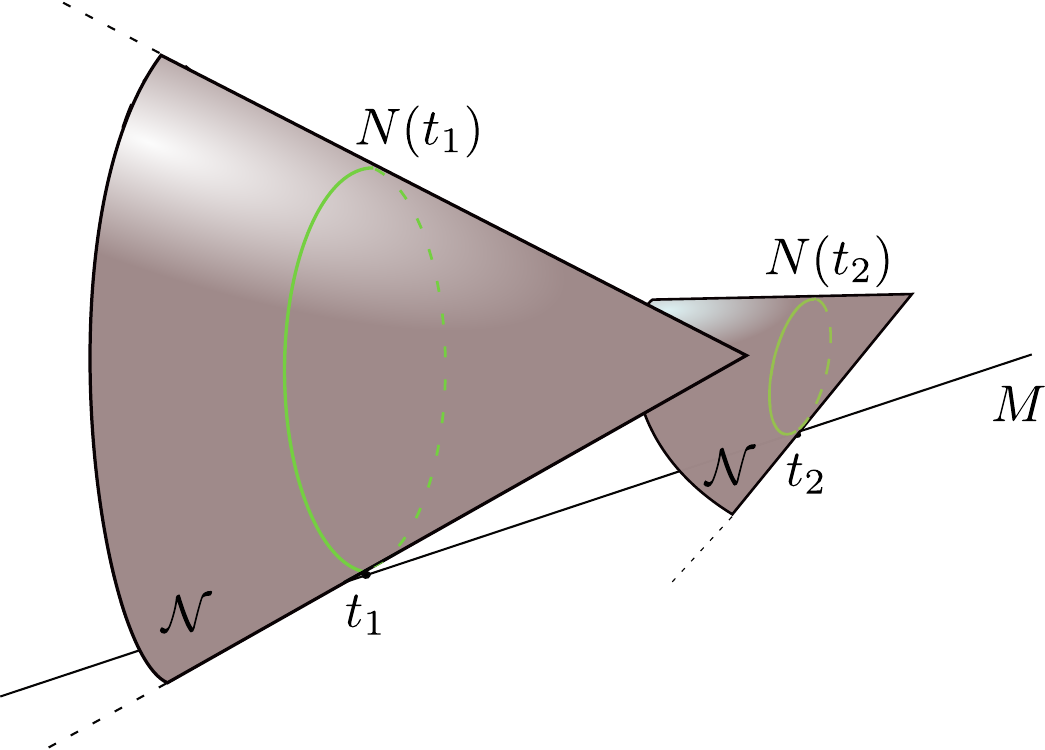}
    \caption{Shape evolution of $N$ when transported on the one dimensional manifold $M$ interpreted as transport of a higher dimensional surface ${\cal N}$ on $M$. At moments $t_{1}$ and $t_{2}$, the conic sections defined by the intersection of ${\cal N}$ with a fixed plane containing the manifold $M$ define different shapes and this may be interpreted as non-metricity of the connection transporting ${\cal N}$ along $M$.}
    \label{fig:rollingcone}
\end{figure}

This interpretation of `what is rolled' can be maintained in the generalization wherein the shape is transported including shape evolution: now the shape $N$ is transported according to an $SL(m+1,R)$ transformation, which corresponds to transport of a shape ${\cal N} = SL(m+1,R)/\mathrm{Stab}(\vec{n})$ where $\mathrm{Stab}(n)$ retains an interpretation as the stabilizer of a vector $n^{A}$ normal to $N$ at the point of contact with $M$. For the group $SL(m+1,R)$, it is shown in Appendix \ref{A1} that

\begin{align}
SL(m+1,R)/\mathrm{Stab}(\vec{n}) \cong \mathbb{R}^{m+1}\setminus\{0\}, \label{orbitstar}
\end{align}
which we can interpret in the three dimensional Euclidean embedding picture (i.e. where $M$ and ${\cal N}$ are submanifolds of Euclidean metric signature embedded in a higher dimensional flat Euclidean space) as a $2$-dimensional cone without apex, with the evolving $N$ corresponding to conic sections of ${\cal N}$ which change under transport. This is illustrated schematically in Figure \ref{fig:rollingcone}. However, it is unclear in the higher dimensional case whether an embedding picture exists where ${\cal N}$ is a manifold of fixed geometry and the evolving $N$ are the higher dimensional analogues of conic sections.

\section{Generalization}
\label{generalizedtransport}
We now consider a more general setup: that where the surface $M$ is a $d$-dimensional submanifold of a $d+1$-dimensional flat Euclidean embedding space with metric tensor $\delta_{AB}$ and Cartesian coordinates $X^{A}$ ($A,B,C=1\dots d+1)$ \footnote{For $d \geq 3$ a codimension-one flat embedding may not be possible for a
general $M$: local isometric embedding requires up to $d(d+1)/2$ flat dimensions.
The embedding picture that we discuss should therefore be regarded as motivating the  introduction of variables
$\{H_{AB}, n^A, B^{A}_{\ph{A}B\mu}\}$ and aiding intuition rather than as a construction to be carried out. It will not be necessary to consider the embedding picture when - in later sections - we discuss Cartan-geometric formulations of gravitational theories in four spacetime dimensions.}. We consider transporting another surface $N$ on $M$ - with points on $M$ coordinatized by $x^{a}$ ($a,b,c=1\dots d$) - so that the point of contact of $N$ with $M$ describes a path $x^{a}(t)$ on $M$. The generalization of (\ref{xtran}) for the evolution of embedding coordinates describing the Cartesian coordinate values of $N$ is given by:

\begin{align}
\tilde{X}^{A}(t_{1}) &=  \Omega^{A}_{\ph{A}B}(t_{1},t_{0})X^{B}(t_{0}) + P^{A}(t_{1},t_{0}) \label{xtrangen}
\end{align}
where now $\Omega^{A}_{\ph{A}B} \in SL(d+1,R)$ with $\Omega^{A}_{\ph{A}B}(t_{0},t_{0})=\delta^{A}_{\ph{A}B}$ and $P^{A}(t_{0},t_{0})=0$. If at $t_{0}$ the surface $N$ is described by an $n$-sphere then the surface consists of points satisfying:

\begin{align}
\delta_{AB}X^{A}X^{B} &= R^{2} \label{initialsphere}
\end{align}
then at $t_{1}$ the surface $N$ is described by points:

\begin{align}
\xi_{AB}(\tilde{X}^{A}-P^{A})(\tilde{X}^{B}-P^{B}) &= R^{2} \label{yxxeq2}
\end{align}
where

\begin{align}
\xi_{AB} \equiv \delta_{CD}(\Omega^{-1})^{C}_{\ph{C}A}(\Omega^{-1})^{D}_{\ph{D}B}
\end{align}
As in the case of one dimensional $M$, we may write (\ref{yxxeq2}) as $F(\tilde{X}^{A})=R^{2}$ and hence the vector normal to the surface at some point of contact $X_{(0)}^{A}$ will be aligned with $\tilde{\nabla}^{A}F =2\xi^{A}_{\ph{A}B}(\tilde{X}^{B}-P^{B})$. Denoting the normal to $M$ at the point of contact by $m^{A}$, the requirement that $m^{A}$ will be antiparallel with $\tilde{\nabla}^{A}F$ implies that

\begin{align}
P^{A} &= \tilde{X}^{A}_{(0)}+\frac{R}{\sqrt{(\xi^{-1})_{CD}m^{C}m^{D}}}(\xi^{-1})^{A}_{\ph{A}B}m^{B}.
\end{align}
As $\Omega^{A}_{\ph{A}B}$ depends on $t$ and not $X^{A}$, the resulting surface is a quadric and describes the surface of a $d$-ellipsoid.


In transporting $N$ from $x^{a}$ to $x^{a}+\delta x^{a}$, the \emph{transported} normal vector $n^{A}$ to $N$ at $x^{a}$ differs from the normal vector to $N$ at $x^{a}+\delta x^{a}$ by an amount $\Delta n^{A}$ defined as:

\begin{align}
\Delta n^{A} \equiv D_{a}n^{A}\delta x^{a} \label{evo1}
\end{align}
where 

\begin{align}
D_{a}n^{A} \equiv \partial_{a}n^{A} + B^{A}_{\ph{A}B a}n^{B}
\end{align}
describes the combined effect of $n^{A}$ varying due to $N$ being in contact with $M$ (with $M$ potentially having extrinsic curvature) in addition to transport - considered as some general combination of shear and rolling -  represented by the effect of $B^{A}_{\ph{A}Ba}$, which is now taken to be in the Lie algebra of $SL(d+1,R)$.

From the perspective of $N$, the transport process inscribes an infinitesimal arc spanning from an original coordinate value $q^{\alpha}$ on $N$ to $q^{\alpha}+\delta q^{\alpha}$ and we should also have:

\begin{align}
\Delta n^{A} &= \frac{\partial n^{A}}{\partial q^{\alpha}}\delta q^{\alpha} \label{evo2}
\end{align}
In transport without slipping we should have a match between the distance rolled on $N$ and on $M$ of an infinitesimal interval of path. This corresponds to a matching between the squared arc length $\delta s^{2}$ inscribed on $N$ by rolling and the squared arc length $g_{ab}\delta x^{a}\delta x^{b}$ along an infinitesimal path element on $M$:

\begin{align}
    \delta s^{2} &= g_{ab}\delta x^{a}\delta x^{b}
\end{align}
We can define the tangent displacement $\delta X^{A}$ (components of a vector in $\mathbb{R}^{d+1}$) as follows:

\begin{align}
\Delta X^{A} &= \frac{\partial X^{A}}{\partial q^{\alpha}}\delta q^{\alpha} \label{um}
\end{align}
Then we should have:

\begin{align}
\delta s^{2} &=  \delta_{AB}\Delta X^{A}\Delta X^{B}\label{dolph}
\end{align}
Now, from the Weingarten equation we have:

\begin{align}
    \frac{\partial n^{A}}{\partial q^{\alpha}}  \delta q^{\alpha}&=  -k_{\alpha}^{\ph{\alpha}\beta}\frac{\partial X^{A}}{\partial q^{\beta}} \delta q^{\alpha}
\end{align}
We can look to formally invert (\ref{um}) as follows:

\begin{align}
\delta q^{\alpha} &=  Q^{\alpha}_{\ph{\alpha}A}\Delta X^{A} \label{ur2}
\end{align}
where $Q^{\alpha}_{\ph{\alpha}A}$ is the Moore-Penrose pseudo-inverse of $\partial X^{A}/\partial q^{\beta}$. Using (\ref{evo1}), (\ref{evo2}), and (\ref{ur2}) we have:

\begin{align}
   D_{a}n^{A}\delta x^{a}&=  -k_{\alpha}^{\ph{\alpha}\beta}\frac{\partial X^{A}}{\partial q^{\beta}} Q^{\alpha}_{\ph{\alpha}B} \Delta X^{B} \\
   & \equiv {\cal Y}^{A}_{\ph{A}B}\Delta X^{B} \label{eq7}
\end{align}
Note that ${\cal Y}^{A}_{\ph{A}B}n_{A}=0$ and that one may locally choose a frame in which ${\cal Y}_{AB} \overset{*}{=} -\text{diag}(\kappa_{1},\dots,\kappa_{d},0)$ where $\kappa_{i}$ are the principal curvatures of $N$\footnote{The notation $\overset{*}{=}$ means equality in a specific frame or gauge}. This tensor can be formally inverted to yield the following:

\begin{align}
    \Delta X^{A} &=  ({\cal Y}^{-1})^{A}_{\ph{A}B}D_{a}n^{B}\delta x^{a}
\end{align}
where $({\cal Y}^{-1})^{A}_{\ph{A}B}$ is the Moore-Penrose pseudo-inverse of ${\cal Y}^{A}_{\ph{A}B}$. So from (\ref{dolph}) we have:

\begin{align}
\delta s^{2} &= \delta_{AB}\Delta X^{A} \Delta X^{B} \nn\\
&\equiv h_{AB} D_{a}n^{A}D_{b}n^{B} \delta x^{a}\delta x^{b}
\end{align}
where: 

\begin{align}
h_{AB} &=  \delta_{CD}({\cal Y}^{-1})^{C}_{\ph{C}A}({\cal Y}^{-1})^{D}_{\ph{D}B} \label{shapetensor}
\end{align}
Note that $h_{AB}n^{A}=0$ and that one may locally choose a frame in which $h_{AB} \overset{*}{=} \text{diag}(1/\kappa_{1}^{2},\dots, 1/\kappa_{d}^{2},0)$. Therefore we identify the metric tensor $g_{ab}$ on $M$ as:

\begin{align}
g_{ab} = h_{AB} D_{a}n^{A} D_{b}n^{B} \label{ghnA}
\end{align}
This is the generalization of equation (\ref{noslip}). We see then that the metric $g_{ab}$ can be recovered from a combination of $\{h_{AB},n^{A},B^{A}_{\ph{A}Ba}\}$. In a general sense, rolling/transport without slipping is generated by the parts of $B^{A}_{\ph{A}Ba}$ that cause $n^{A}$ to change during the transport (for example, in the pure rotation case in the previous section, the normal vector to the surface of the circle at point $p_{0}$ gets rotated as the circle rolls to another point $p_{1}$). 

Note that $g_{ab}$ is unchanged if in equation (\ref{ghnA}) in place of $h_{AB}$ one uses:

\begin{align}
    H_{AB} &=  h_{AB} + \alpha n_{A}n_{B} \label{HABdef}
\end{align}
for constant $\alpha$. It will turn out to be useful to use $H_{AB}$ in later sections when considering these quantities as degrees of freedom in gravitational theories.

\subsection{The scope for further generalization}

We note that two shapes which have the same second fundamental forms along the path of contact points therefore yield identical
transport-without-slipping data, and it suffices to follow $N$ at the contact point which, where $N$ is locally convex, is an
ellipsoid and so lies in the $GL(d+1,\mathbb{R})$ orbit of the sphere (\ref{initialsphere}). The
assumption that $\Omega^{A}_{\ph{A}B}$ is independent of $X^A$ is therefore not a restriction as
far as the recovery of $g_{ab}$ is concerned, although it is essential to the
Cartan-geometric interpretation, since an $X^{A}$-dependent $\Omega^{A}_{\ph{A}B}$ would not yield
a $B^{A}_{\ph{A}Ba}$ valued in a finite-dimensional Lie algebra. 

A natural starting point for relaxing the assumption that $\Omega^{A}{}_{B}$ is independent
of $X^{A}$ is to take as the arena not a finite-dimensional group orbit but the space of
shapes itself. Labelling each deformation of the $n$-sphere by its radial distance from a
chosen centre in each direction, this space may be presented as
\begin{equation}
\mathcal{N} = \Big( \prod_{s \in S^{n}} \mathbb{R}^{1}_{+} \Big) \big\backslash\, \sigma ,
\end{equation}
where $\sigma$ denotes the subset of $\prod_{s\in S^n}\mathbb{R}_+^1$ where there appear any non-smooth elements amongst the shape's points, i.e. the axes representing the distance from the centre do not form an everywhere smooth shape.
A shape evolution is then a curve in $\mathcal{N}$, and the data
relevant to transport is a curve in the associated space
$\Delta = \bigcup_{N \in \mathcal{N}} \{N\} \times N$ of pairs (shape, point on that shape),
whose projection to the embedding space is required to follow the path $x^{a}(t)$ on $M$.
The shapes considered in Section 3 -- the $SL(d+1,\mathbb{R})$ orbit of the sphere (44) --
constitute a finite-dimensional submanifold of $\mathcal{N}$, and the tangent directions to
$\mathcal{N}$ transverse to it are precisely those deformations that vary from point to
point on $N$. Modulo tangential reparametrizations, a tangent vector to $\mathcal{N}$ is a
single normal deformation function on the shape.

The requirement of
transport without slipping does not appear to obstruct this generalization: for a path on
$M$ and a prescribed evolution of $N$, one expects the existence of a contact path along
which the tangent spaces remain aligned, the local analysis proceeding as in Section 3 with
the soldering form of $N_{0}$ composed with the differential of the deformation map.
Moreover, the recovery of $g_{ab}$ depends only on the second fundamental form of $N$ along
the path of contact points, so a large class of deformations (for example any supported away from the
contact point) is invisible to the metric and is available to carry other information.  It is an open question as to whether a useful notion of curvature, and hence
of encoded geometry,  survives in this setting. Additionally, an interesting question  is what the new shape degrees of freedom on $N$ would correspond to on $M$. 

\subsection{The existence of transport without slipping}

Transport
without slipping can always be arranged locally: since the orbit of $n^A$ is open
in $\mathbb{R}^{d+1}$ by (\ref{orbitstar}), the quantity $D_an^A$ may be prescribed freely, and
(\ref{ghnA}) then requires only that its tangential part $E^I_a$ be a linear isometry from
$(T_xM,g_{ab})$ to the tangent space of $N$ carrying $h_{IJ}$. Such a map exists
whenever the two-forms are non-degenerate and of matching signature, and is unique
up to an $O(d)$ frame rotation. We emphasize that these results are local in the parameters
along the path and there may exist important restrictions when considering finite paths. For example, as was seen in Section \ref{sheartransport}, it was possible to transport an ellipse along a line entirely by shear evolution but only for a finite path along the line. It remains an open question as to whether transport without slipping can be achieved for a finite path of a surface with shape $N_{1}$ at the beginning of the path and $N_{2}$ at the end. The process of transport without slipping
therefore fixes the soldering form up to a frame rotation and leaves the components of $B^{A}_{\ph{A}Ba}$ that do not generate infinitesimal changes in $n^A$ entirely free. The remainder of this article will focus on the different ways in which this unfixed part of $B^{A}_{\ph{A}Ba}$ may encode further information about the geometry of $M$.

\begin{figure}[H]
    \centering
    \includegraphics[width=0.65\linewidth]{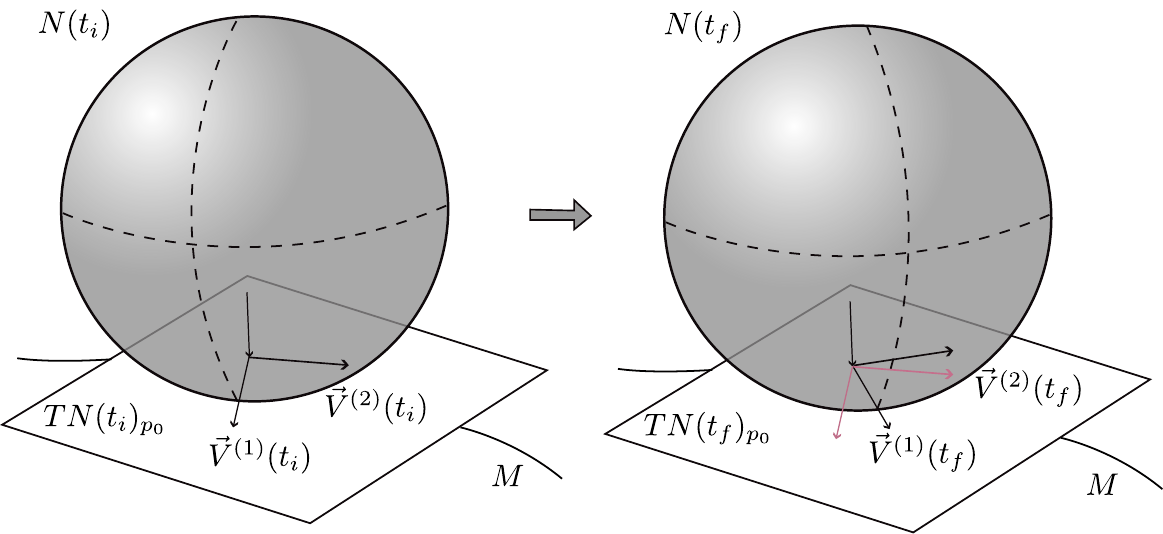}
    \caption{Rotation of sphere after rolling without slipping or twisting around a small closed loop on the manifold $M$}
    \label{fig:standardcartan}
\end{figure}

By way of example, consider rolling a sphere of fixed volume on a two dimensional manifold $M$ embedded in $\mathbb{R}^{3}$. Then, for two points $t_{1}$ and $t_{0}$ on a curve $x^{a}(t)$ on $M$, the points on the surface of the sphere in terms of embedding coordinates $X^{A}$ of the sphere's surface, will evolve as follows:

\begin{align}
\tilde{X}^{A}(t_{1}) &=  {\cal R}^{A}_{\ph{A}B}(t_{1},t_{0})X^{B}(t_{0}) + P^{A}(t_{1},t_{0}) \label{xtran2}
\end{align}
with ${\cal R}^{A}_{\ph{A}B} \in SO(3)$. If we consider a point $t_{1}=  t_{0}+\delta t$ located at a point $x(t_{0}) + \delta x$ on the path, we have that

\begin{align}
    {\cal R}^{A}_{\ph{A}B} = \delta^{A}_{\ph{A}B} - A^{A}_{\ph{A}B a}\delta x^{a}  + \dots {\cal O}(\delta x^{2})
 \end{align}
where each $A^{A}_{\ph{A}B a}\delta x^{a}$ is in the Lie algebra of $SO(3)$ and may be written as $\alpha_{(I)}T^{(I)A}_{\ph{(I)A}B}$ where $T^{(I)A}_{\ph{(I)A}B}$ are the generators of $SO(3)$. For a given $\delta x^{a}$, two of these generators would change the point of contact of the sphere $N$ with $M$, one of which would generate a change of normal vector perpendicular to the direction of $\delta x^{a}$ and so would generate slipping. The remaining generator does not change the point of contact. In typical treatments, it is constrained so that vectors in the tangent spaces of $M$ and $N$ are parallel transported according to the Levi-Civita connection of their respective metrics - this corresponds to rolling the ball \emph{without twisting}. This leads to a remarkable result - that when a vector in the embedding space, initially in the tangent plane of $N$ at $t_{0}$ at the point of contact, is rolled around a small loop according to the entire $SO(3)$ connection $A^{A}_{\ph{A}Ba}$, then upon returning to $t_{0}$ the vector will have rotated in the tangent plane of $N$ - represented by an $SO(2)$ subgroup of the $SO(3)$ rolling group - by an amount dictated by the difference in Riemannian curvature between $M$ and $N$ (see Figure \ref{fig:standardcartan}) \cite{Wise:2006sm}.

\section{Application to gravitation and spacetime: MacDowell-Mansouri gravity}
\label{section-macdowellmansouri}

Heuristically, this approach can be extended to the geometry of spacetime, where both $M$ and $N$ are four-dimensional and have Lorentzian signature metrics. For example if $N$ is de Sitter space then the group of `rotations' in $TN_{p}$ is the group $SO(1,3)$ which is taken to be a subgroup of the group $SO(1,4)$ which can be regarded as the set of transformations that `roll' the de Sitter space in $\mathbb{R}^{(1,4)}$. The basic variables necessary to define the metric are then a `contact vector' $n^{A}$ and a rolling connection $A^{A}_{\ph{A}\mu}$. Given that the shape $N$ is assumed to be fixed during the rolling process, then the - now $SO(1,4)$ -  tensor $H_{AB}$ of (\ref{HABdef}) can be taken to be simply proportional to the $SO(1,4)$-invariant matrix $\eta_{AB}=\mathrm{diag}(-1,1,1,1,1)$: $H_{AB} = \ell^{2} \eta_{AB}$, where $\ell$ is the radius of the de Sitter space $N$ and hence the metric of spacetime can be defined to be:

\begin{align}
g_{\mu\nu} &= H_{AB}D_{\mu}n^{A} D_{\nu}n^{B} = \ell^{2} \eta_{AB}D_{\mu}n^{A} D_{\nu}n^{B}
\end{align}
where $D$ is the $SO(1,4)$-covariant exterior derivative. Action principles for gravity with $n^{A}$ and $A^{A}_{\ph{A}B\mu}$ as basic variables were first considered in \cite{Stelle:1979aj} (there in the context of the group $SO(2,3)$, but the results are easily adapted to the present case), based on earlier results \cite{MacDowell:1977jt}. For example:

\begin{align}
S[A^{A}_{\ph{A}B},n^{A},\lambda] &= -\int \eps_{ABCDE}n^{E}Dn^{A} \wedge Dn^{B} \wedge F^{CD} +   \lambda (\eta_{AB}n^{A}n^{B}-1) \label{actsw}
\end{align}
where the four-form $\lambda$ enforces the fixed norm condition on $n^{A}$. One may enforce this condition at the level of the action and adopt a partial gauge fixing condition $n^{A} \overset{*}{=} -\delta^{A}_{\ph{A}4}$. In this gauge we have:

\begin{align}
g_{\mu\nu} &= \ell^{2}\eta_{IJ}A^{I}_{\ph{I}4\mu}A^{J}_{\ph{J}4\nu}
\end{align}
where $\eta_{IJ}=\mathrm{diag}(-1,1,1,1)$ is the invariant matrix of $SO(1,3)$. Comparing this expression to (\ref{geqee}) suggests the identification of $A^{I}_{\ph{I}4\mu}$ as being proportional to the co-tetrad $e^{I}_{\mu}$ and hence in this gauge we can write the $SO(1,4)$ connection as:

\begin{align}
    A^{AB}_{\ph{AB}a} &\overset{*}{=} \begin{pmatrix} 
   \omega^{IJ}_{\ph{IJ}a} & - \frac{1}{\ell}e^{I}_{a} \\
  \frac{1}{\ell}  e^{I}_{a} & 0
    \end{pmatrix}
\end{align}
where $\omega^{IJ}_{\ph{a}}=-\omega^{JI}_{\ph{a}}\in \mathfrak{so(1,3)}$. Then it can be checked that the action (\ref{actsw}) in this gauge takes the form:

\begin{align}
S[\omega^{IJ},e^{I}] &=  \int \eps_{IJKL}e^{I}\wedge e^{J}\wedge \big(R^{KL}-\frac{1}{\ell^{2}}e^{K}\wedge e^{L}\big)
\end{align}
This is the action for Einstein-Cartan gravity with a positive cosmological constant $\Lambda\sim 1/\ell^{2}$. More recently some authors have looked at $SO(1,4)$ or $SO(2,3)$ symmetric actions where $n^{A}$ is unconstrained and its dynamics may have a significant gravitational effect - for example it was shown that an action polynomial in $A^{A}_{\ph{A}B}$ and $n^{A}$ corresponded to a scalar-tensor theory with Peebles-Ratra type potential \cite{Westman:2013mf}.

We will look to generalize this result so that vectors will instead be parallel transported according to a connection $B^{A}_{\ph{A}Ba}$ valued in the Lie algebra of $SL(d+1,R)$ or $GL(d+1,R)$ as description of the rolling/transport process. We will maintain the condition of transport without slipping throughout but will relax the `no twisting' condition. It will be shown that information about the Riemannian curvature of $M$ can -  instead of rotation of a vector in the tangent plane of $N$ - be encoded in `displacement' or tilt of the normal vector to $N$ after rolling around a closed curve on $M$ (which will be interpretable in terms of torsion of the connection) and that this may be identified with teleparallel geometry. The extent to which, alternatively, a specific change of shape of $N$ after being rolled around a small loop (which will be interpretable in terms of non-metricity of the connection) can yield information about the intrinsic curvature of $M$ will furthermore be examined.


\section{Torsion and teleparallelism}
\label{torsiontele}
We consider again the case where a sphere $N$ of radius $\ell$ is rolled on a surface $M$ without slipping. However, we now relax the condition of no-twisting - this means that vectors in the tangent spaces of $N$ and $M$ along the paths on their surface defined by the rolling will be parallel transported according to a connection valued in the Lie algebra of $SO(3)$ which will contain an object more general than the Levi-Civita connection on the respective surfaces.  Choosing an $SO(3)$ `gauge' where $n^{A} \overset{*}{=}(0,0,-1)$ and $h_{AB} \overset{*}{=}\mathrm{diag}(\ell^{2},\ell^{2})$ we have:

\begin{align}
    A^{AB}_{\ph{AB}a} &\overset{*}{=} \begin{pmatrix} 
   \omega^{IJ}_{\ph{IJ}a} & - \frac{1}{\ell}e^{I}_{a} \\
  \frac{1}{\ell}  e^{I}_{a} & 0
    \end{pmatrix}
\end{align}
where $I,J,K,\dots = 1,2$. The part of $A^{AB}_{\ph{AB}a}$ that changes the point of contact is identified with the co-tetrad/soldering form as - following from (\ref{ghnA}):

\begin{align}
g_{ab} &=  \delta_{IJ}e^{I}_{a}e^{J}_{b}
\end{align}
The curvature two-form of the connection $A^{AB}_{\ph{a}}$ is as follows:

\begin{align}
F^{AB}_{\ph{AB}ab} &=  2\partial_{[a}A^{AB}_{\ph{AB}b]} + 2 A^{A}_{\ph{A}C [a}A^{CB}_{\ph{CB}b]} \\
&=  \begin{pmatrix}
R^{IJ}_{\ph{IJ}ab} -\frac{2}{\ell^{2}} e^{I}_{[a}e^{J}_{b]}& -\frac{1}{\ell}T^{I}_{ab} \\
\frac{1}{\ell}T^{I}_{ab} & 0
\end{pmatrix}
\end{align}
where

\begin{align}
R^{IJ}_{\ph{IJ}ab} &= 2\partial_{[a}\omega^{IJ}_{\ph{IJ}b]} + 2 \omega^{I}_{\ph{I}K [a}\omega^{KJ}_{\ph{KJ}b]}  \\
T^{I}_{ab} &=  2\partial_{[a}e^{I}_{b]} + 2 \omega^{I}_{\ph{I}J[a}e^{J}_{b]}
\end{align}
i.e. $R^{IJ}_{\ph{IJ}ab}$ is the curvature two-form of $\omega^{IJ}_{\ph{IJ}a}$ and $T^{I}_{ab}$ is the torsion two-form of $\omega^{IJ}_{\ph{IJ}a}$. To make progress we can decompose $\omega^{IJ}_{\ph{IJ}a}$ as follows:

\begin{align}
\omega^{IJ}_{\ph{IJ}a} &= \bar{\omega}^{IJ}_{\ph{IJ}a} + C^{IJ}_{\ph{IJ}a} \label{spincode}
\end{align}
where $\bar{\omega}^{IJ}_{\ph{IJ}a}(e,\partial e)$ is identified as the torsion-free/Levi-Civita spin connection in the Lie algebra of $SO(2)$ \footnote{This is the solution to the equation $de^{I}+\omega^{I}_{\ph{I}J}\wedge e^{J}=0$.}. As such, $C^{IJ}_{\ph{IJ}a}$ is identified with the contorsion one-form and encodes the deviation away from Levi-Civita transport of vectors transported along the curve $x^{a}(t)$ on $M$. The physical interpretation of this is an additional twisting of the ball as it is rolled along the curve.

\begin{figure}[H]
    \centering
    \includegraphics[width=0.65\linewidth]{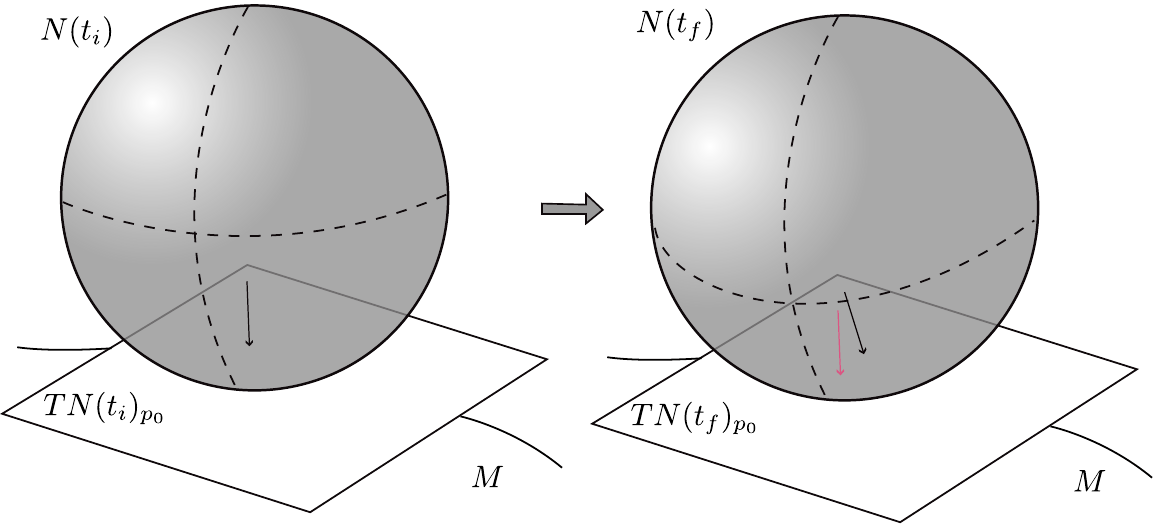}
    \caption{The change of $N$ under rolling without slipping but with torsionful twisting around a small loop on $M$. After transport around the loop, the ball has tilted and the transported normal vector to the surface no longer aligns with the original normal vector at $p_{0}$. This implies that a closed rolled path on $M$ will correspond to a non-closed rolled path inscribed on $N$, as shown in Figure \ref{fig:torsionpath}}
    \label{fig:torsionroll}
\end{figure}

\begin{figure}[H]
    \centering
    \includegraphics[width=0.5\linewidth]{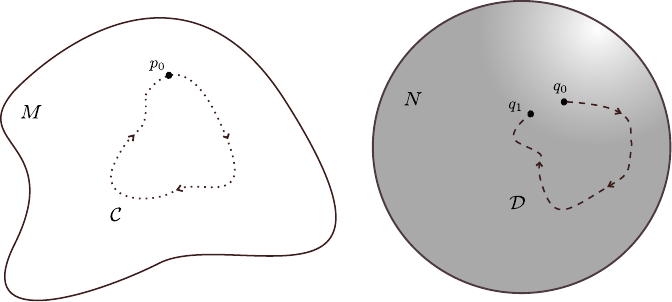}
    \caption{Torsionful transport of $N$ on $M$. The ball $N$ is rolled without slipping but with twisting on $M$. Though paths ${\cal C}$ and ${\cal D}$ have the same length according to the embedding space metric, the closed path ${\cal C}$ begins and ends at $p_{0}$ whilst the corresponding path ${\cal D}$ inscribed on $N$ begins at a point $q_{0}$ and finishes at a different point $q_{1}$. The difference between $q_{1}$ and $q_{0}$ may be interpreted in terms of torsion of the rolling connection.}
    \label{fig:torsionpath}
\end{figure}

An important result can be obtained by considering the rolled path parameterized by $t\in [t_{i},t_{f}]$ to consist of an infinitesimal parallelogram spanned by vectors $\vec{\delta x} = \delta x \partial/\partial x$ and $\vec{\delta y} = \delta y \partial/\partial y$. Then we may compare a vector at $p_{0}$ before and after rolling around this path. For a vector $U^{I}$ in the tangent plane of $TN(t_{i})$, we have \cite{Westman:2014yca} that 

\begin{align}
\Delta U^{I} &= U^{I}(t_{f}) - U^{I}(t_{i}) \\
&= \bigg(R^{IJ}_{\ph{IJ}ab} -\frac{2}{\ell^{2}} e^{I}_{[a}e^{J}_{b]}\bigg) \delta x^{a}\delta y^{b} U^{J} \label{RIJU}
\end{align}
where the right hand side of (\ref{RIJU}) is evaluated at $t_{0}$ and higher order terms in $(\delta x,\delta y)$ have been assumed to be negligible. Additionally, it can be shown that the normal vector $n^{A}$, after rolling around the closed path, will change by an amount as follows:

\begin{align}
\Delta n^{A} &= n^{A}(t_{f}) - n^{A}(t_{i}) \\
&= \frac{1}{\ell}T^{A}_{\ph{A}ab}\delta x^{a}\delta y^{b}
\end{align}
Note that to this order, $n_{A}\Delta n^{A}=0$ due to $SO(3)$ transport preserving the length $n_{A}n^{A}$ of vectors. This implies that $n_{A}T^{A}_{\ph{A}ab}=0$. Therefore the effect of torsion is seen as a specific rotation of the sphere after transport of the sphere around the closed loop such that the normal vector `tilts' off the surface (see Figure \ref{fig:torsionroll}). Interpreted in terms of the rolling of a sphere $N$ on a manifold $M$, a non-zero $\Delta n^{A}$ will correspond to a non-closed path inscribed on $N$ after rolling around a closed loop on $M$ (see Figure \ref{fig:torsionpath}); this is a Cartan-geometrical version of the statement of `parallelograms not closing' as an interpretation of torsionful transport. 

Given the decomposition (\ref{spincode}) , we can decompose the forms $R^{IJ}_{\ph{IJ}ab}$ and $T^{I}_{ab}$ as follows:

\begin{align}
R^{IJ}_{\ph{IJ}ab} &= \bar{R}^{IJ}_{\ph{IJ}ab}(\bar{\omega})+ 2D^{(\bar{\omega})}_{[a}C^{IJ}_{b]} +2 C^{I}_{\ph{I}K [a}C^{KJ}_{\ph{KJ}b]}\label{rijdeco}\\
 T^{I}_{ab} &= 2C^{I}_{\ph{I}J[a}e^{J}_{b]}
\end{align}
where $\bar{R}^{IJ}_{\ph{IJ}ab}(\bar{\omega})$ is the curvature two-form of the torsion-free spin connection $\bar{\omega}^{IJ}_{\ph{IJ}a}$ and $D_{a}^{(\omega)}$ is the covariant derivative associated with it.

In the teleparallel geometry it is required that the curvature $R^{IJ}_{\ph{IJ}ab}$ is itself zero.\footnote{See \cite{Barker:2022jsh} for a generalization of this notion to \emph{partial teleparallelism}, wherein the wider case of possible constraints within the set of irreducible parts of torsion and curvature two-forms is considered.}  Clearly from (\ref{rijdeco}) this will only generally be possible for a specific, non-vanishing form of $C^{I}_{\ph{I}Ja}$. In this sense, teleparallelism replaces the curvature of a spin connection with its torsion as a measure of the intrinsic curvature of $M$.

Explicitly, the curvature $R^{IJ}_{\ph{IJ}ab}$ will vanish if \cite{Eisenhart1927}:

\begin{align}
\omega^{IJ}_{\ph{IJ}a} &= (\Lambda^{-1})^{I}_{\ph{I}K}\partial_{a}(\Lambda^{KJ})
\end{align}
where $\Lambda^{I}_{\ph{I}J} \in SO(2)$ so from (\ref{spincode}) we would require:

\begin{align}
C^{IJ}_{\ph{IJ}a} &= (\Lambda^{-1})^{I}_{\ph{I}K}\partial_{a}(\Lambda^{KJ}) - \bar{\omega}^{IJ}_{\ph{IJ}a}
\end{align}
From the perspective of Cartan geometry, an alternative useful definition of teleparallelism would be if the $I,J$ components $F^{IJ}_{\ph{IJ}a}$ of the entire $SO(3)$ rolling connection vanished, which would mean - via (\ref{RIJU}) a replacement of rotation of tangent vectors at the point of contact after transport around a loop with a `tilt' of the sphere after such transport.\footnote{See \cite{Baez:2012bn} for an alternate perspective on teleparallelism within a formalism termed Cartan 2-geometry.} This would be the case if $R^{IJ}_{\ph{IJ}ab}$ was equal to the curvature two-form of a sphere of radius $\ell$, with corresponding Levi-Civita spin connection $\omega^{(S)IJ}_{\ph{(S)IJ}a}$ and so the requirement on contorsion would then be:

\begin{align}
C^{IJ}_{\ph{IJ}a} &= \omega^{(S)IJ}_{\ph{(S)IJ}a} - \bar{\omega}^{IJ}_{\ph{IJ}a}
\end{align}
Applying this formalism to the case of spacetime torsion as an alternative to Riemannian curvature i.e. the teleparallel formalism, the appropriate gauge symmetry is $SO(1,4)$ or $SO(2,3)$. We will focus on the former case, but the formalism can easily be adapted to the latter. 

Making use of the $SO(1,4)$-invariant tensor $\eta_{AB}=\mathrm{diag}(-1,1,1,1,1)$ we can consider $n^{A}$ to be constrained to be of unit-norm with respect to $\eta_{AB}$:

\begin{align}
\eta_{AB}n^{A}n^{B} = 1
\end{align}
Then we may define a projector

\begin{align}
P^{A}_{\ph{A}B} &= \delta^{A}_{\ph{A}B}+ n^{A}n_{B}
\end{align}
where $n_{A} = \eta_{AB}n^{B}$ and $P^{A}_{\ph{A}B}n^{B}=0$ so we have that 
\begin{align}
T^{A} &= \ell F^{A}_{\ph{A}B}n^{B}\\
R^{AB} &= P^{A}_{\ph{A}C}P^{B}_{\ph{B}D}F^{CD} + \frac{1}{\ell^{2}} Dn^{A} \wedge Dn^{B}
\end{align}
where $T^{A}n_{A}=R^{AB}n_{B}=0$.
Then, for example, we may identify the most general tensor quadratic in terms of projections of the full $SO(1,4)$ or $SO(2,3)$ curvature two-form along $n^{A}$

\begin{align}
F^{A}_{\ph{A}BCD}F^{E}_{\ph{E}GHI}n^{B}n^{I} = \frac{1}{\ell^{2}}T^{A}_{\ph{A}CD}T^{E}_{\ph{E}GH} \label{FintoT}
\end{align}
and then $e_{a}^{I} = D_{a}n^{I}$ can be used to change internal indices into spacetime coordinate indices as necessary. Then, an $SO(1,4)$-symmetric action principle for teleparallel gravity would be:

\begin{align}
S[A^{A}_{\ph{A}B},n^{A},\lambda^{A}_{\ph{A}B},\lambda] =  -\frac{\ell^{2}}{16\pi G} &\int d^{4}x\sqrt{-g}\bigg(\frac{1}{4}F_{ABCD}F^{AECD}+\frac{1}{2}F_{ABCD}F^{CEAD}-F^{A}_{\ph{A}BCA}F^{DEC}_{\ph{DEC}D}\bigg)n^{B}n_{E}\nn\\
+&\int \lambda_{AB}\wedge\bigg(P^{A}_{\ph{A}C}P^{B}_{\ph{B}D}F^{CD} + \frac{1}{\ell^{2}} Dn^{A} \wedge Dn^{B}\bigg) +\lambda(\eta_{AB}n^{A}n^{B}-1)
\label{telact}
\end{align}
where the Lagrange multiplier fields $\lambda^{A}_{\ph{A}B}$ and $\lambda$ enforce the vanishing of $R^{AB}$ and the fixed norm condition on $n^{A}$ respectively. Hence, we can see that via (\ref{FintoT}), the specific combination of curvature squared terms appearing in (\ref{telact}) may be expressed in terms of quadratic invariants of the torsion and it can be checked that this specific combination of them yields the teleparallel equivalent of General Relativity \cite{Aldrovandi:2013wha}.

\section{Non-metricity}
\label{section-nonmetricity}

We have seen then that the probing of geometry by rolling a sphere without slipping or twisting can alternatively be done by rolling without slipping but with a specific recipe for twisting. We now explore a different possibility - that of using evolution of the actual \emph{shape} of $N$ to probe the geometry of $M$. To this end, we consider the setup of Section \ref{generalizedtransport} with transport of $N$ which at an initial point on the rolled path $x^{a}(t_{0})$ is assumed to be a two-sphere but which can evolve into an ellipsoid at later points on the path. Rolling of vectors such as $n^{A}$ will now be done by a connection valued in the Lie algebra of $SL(3,R)$. In a gauge where $n^{A}=(0,0,1)$, the transport/rolling connection $B^{A}_{\ph{A}Ba}$ can be decomposed as:

\begin{align}
B^{A}_{\ph{A}Ba} &\overset{*}{=}  \begin{pmatrix}
-\frac{1}{2}c_{a}\delta^{I}_{\ph{I}J} + w^{I}_{\ph{I}Ja } & E^{I}_{a} \\
V_{Ia} & c_{a}
\end{pmatrix}  \label{bcon}
\end{align}
where $w^{I}_{\ph{I}Ja}$ is valued in the Lie algebra of $SL(2,R)$. 
From (\ref{ghnA}) and (\ref{HABdef}) we have:

\begin{align}
g_{ab} &=H_{AB}D_{a}n^{A}D_{b}n^{B} \overset{*}{=} h_{IJ}E^{I}_{a}E^{J}_{b}
\end{align}
Recall that by a suitable orthogonal transformation at a point $h_{IJ}\rightarrow \lambda^{K}_{\ph{K}I}\lambda^{L}_{\ph{L}J}h_{KL}$, we have $h_{IJ} = \mathrm{diag}(\kappa_{1}^{-2},\kappa_{2}^{-2})$, where $\kappa_{(I)}$ are the principal curvatures of $N$. We can then relate the co-tetrad to $E^{I}_{a}$ via

\begin{align}
E^{I}_{a} &= {\cal Y}^{I}_{\ph{I}J}e^{J}_{a} \label{eia}
\end{align}
where ${\cal Y}^{I}_{\ph{I}J}$ are $I,J$ components of the tensor ${\cal Y}^{A}_{\ph{A}B}$ defined in (\ref{eq7}). 


Additionally, the components $h_{IJ}$ of the `shape tensor' $h_{AB}$ (\ref{shapetensor}) - and the associated inverse matrix $h^{IJ}$ - can be used to raise and lower indices of $(w^{I}_{\ph{I}Ja},E^{I}_{a},V_{Ia})$ so for example $w^{IJ}_{\ph{IJ}a}$ is understood to be $w^{I}_{\ph{I}Ka}h^{KJ}$. We furthermore decompose:

\begin{align}
w^{IJ}_{\ph{IJ}a} &=  A^{IJ}_{\ph{IJ}a} + S^{IJ}_{\ph{IJ}a}
\end{align}
where $A^{(IJ)}_{\ph{(IJ)}a}=0$ and $S^{[IJ]}_{\ph{[IJ]}a}=0$ and $S^{I}_{\ph{I}Ia}=0$. As such, we can interpret $A^{IJ}_{\ph{IJ}a}$ as generating rotations and $S^{IJ}_{\ph{IJ}a}$ as generating shears of the transported shape. The curvature two-form can then be decomposed as:

\begin{align}
F^{I}_{\ph{I}Jab} &=  2\partial_{[a}A^{I}_{\ph{I}J|b} + 2A^{I}_{\ph{I}K[a}A^{K}_{J|b]} + 2S^{I}_{\ph{I}K[a}S^{K}_{J|b]}
+ 2\partial_{[a}S^{I}_{\ph{I}J|b]} + 2A^{I}_{\ph{I}K[a}S^{K}_{J|b]} + 2S^{I}_{\ph{I}K[a}A^{K}_{J|b]} \label{fij}\nn\\
&+ 2E^{I}_{[a} V_{J|b]}
- \partial_{[a}c_{b]}\delta^{I}_{\ph{I}J}\\
F^{I}_{\ph{I}3ab} &=
2\partial_{[a}E^{I}_{b]} + 2(A^{I}_{\ph{I}K[a}+ S^{I}_{\ph{I}K[a}) E_{b]}^{K}
-3 c_{[a}E^{I}_{b]} \label{curvature2} \\
F^{3}_{\ph{3}Jab} &=
2\partial_{[a}V_{J|b]} - 2(A^{K}_{\ph{J}J[a} + S^{K}_{\ph{K}J[a})V_{K|b]}
+ 3c_{[a} V_{J|b]} \label{curvature3}\\
F^{3}_{\ph{3}3ab} &= 
2\partial_{[a}c_{b]} + 2V_{K[a} E^{K}_{b]} \label{curvature4}
\end{align}




Again considering parallel transport around a small closed path we have that the quantity $h_{IJ}$ will change by an amount as follows:

\begin{align}
\Delta h_{IJ} &=  h_{IJ}(t_{f}) - h_{IJ}(t_{i}) \\
&= (-F^{K}_{\ph{K}Iab}h_{KJ}-F^{K}_{\ph{K}Jab}h_{IK})\delta x^{a}\delta y^{b} \\
&= -2F_{(IJ)ab}\delta x^{a}\delta y^{b} \label{hijholo}
\end{align}
This quantity is distinct from what is typically defined as the \emph{non-metricity} form $Q_{ABa}$ defined as follows:

\begin{align}
    Q_{ABa} &=  -D_{a}h_{AB} = -\partial_{a}h_{AB} + B^{C}_{\ph{C}Aa}h_{CB} + B^{C}_{\ph{C}Ba}h_{AC}  \label{nonmetform}
\end{align}
where $D= d + B$. However it is strictly tied to non-metricity and vanishes if non-metricity is zero. Indeed, it is not so difficult to show that the quantity appearing in (\ref{hijholo}) is just the field strength of non-metricity (see for instance \cite{Iosifidis:2019dua,BeltranJimenez:2020sih}):

\begin{equation}
    F_{(IJ)}=\frac{1}{2}D Q_{IJ}
\end{equation}
In the torsionful teleparallel scenario discussed in Section \ref{torsiontele}, the geometry of $M$ was essentially encoded in the contorsion form $C^{IJ}_{\ph{IJ}a}$. The extension of the transport/rolling process to include shears allows the introduction of a new form: the non-metricity form $Q_{ABa}$ (\ref{nonmetform}). For the remainder of this section we will focus on the degree to which geometry can instead be encoded in $Q_{ABa}$. 

\subsection{Analogue to symmetric teleparallelism}
\label{subsection-symtele}

We now show that this is possible in a scenario that is closely analogous to the
symmetric teleparallelism (STP) \cite{BeltranJimenez:2019esp} formalism. Firstly we assume a form of the connection $B^{A}_{\ph{A}Ba}$ which allows the  gauge in which $V^{I}_{a} \overset{*}{=} 0$ to exist. Additionally we assume that $\partial_{[a}c_{b]}=0$ and so - locally -we may further move to a gauge $c_{a}\overset{*}{=} 0$, in which case the curvature two-form takes the following form:

\begin{align}
F^{A}_{\ph{A}Bab} &\overset{STP}{=} \begin{pmatrix}
  R^{I}_{\ph{I}Jab}(w) &  2\partial_{[a}E^{I}_{b]}+ 2w^{I}_{\ph{I}J[a} E_{b]}^{J} \\
0 & 0
\end{pmatrix}
\end{align}
It is then required that the connection $w^{I}_{\ph{I}Ja}$ has vanishing curvature two-form $R^{IJ}_{\ph{IJ}ab}=  0$ which implies at least locally we have that

\begin{align}
w^{I}_{\ph{I}Ja} &\overset{*}{=} (\Sigma^{-1})^{I}_{\ph{I}K}\partial_{a}(\Sigma)^{K}_{\ph{K}J}
\end{align}
where $\Sigma^{I}_{\ph{I}J} \in SL(2,R)$. Note that this implies that both $F^{(IJ)}_{\ph{(IJ)}ab}$ and $F^{[IJ]}_{\ph{[IJ]}ab}$ are zero. We further impose that $F^{I}_{\ph{I}3ab} \overset{*}{=} 0$ so that in the gauge $w^{I}_{\ph{I}Ja} \overset{*}{=} 0$ we have that $E_{a}^{I} \overset{*}{=} \partial_{a}\xi^{I}$. Additionally in this gauge we have that the non-metricity is given by:

\begin{align}
Q_{ABa} =  -D^{(w)}_{a}h_{AB} &\overset{*}{=} -\partial_{a}h_{AB} 
\end{align}
and - from (\ref{ghnA}) - and recalling that we have assumed the existence of a gauge in which $n^{A} = \delta^{A}_{\ph{A}3}$ then:

\begin{align}
g_{ab} &\overset{*}{=} h_{IJ} \partial_{a}\xi^{I}\partial_{b}\xi^{J}
\end{align}
As $F^{IJ}_{\ph{IJ}ab}=0$, it follows from (\ref{hijholo}) that $\Delta h_{IJ}$ will be equal to zero in this case, and hence no change in shape after transport around a small closed loop. However, the condition $V_{Ia}=0$ is a higher dimensional analogue of the combined rotation and shear transport of an ellipse discussed in Section \ref{combinedrotationshear} and so a shape that is initially a sphere at $t_{0}$ would evolve into an ellipsoid at later points on a path. 

It is then assumed that $\xi^{I}$ can be interpreted locally as a set of spacetime coordinates so that in the spacetime gauge where they are adopted as coordinates we have:

\begin{align}
h_{IJ} &\overset{*}{=} g_{IJ}\\
Q_{IJK} &\overset{*}{=}-\partial_{K}g_{IJ} 
\end{align}
Then, $Q_{IJK}$ may be used to build an object linear in derivatives of $g_{IJ}$ that essentially plays the role of the Christoffel symbols. In this sense, the geometry can be encoded in the non-metricity which, in the transport/rolling picture indeed corresponds to shape evolution of the rolled shape along a segment of path on the manifold $M$. 

The extension to the geometry of higher dimensional manifolds $M$ then follows: for the case of spacetime, the `rolling/transport' group will be $SL(5,R)$ if it is assumed that the volume of the shape $N$ is preserved by transport, or $GL(5,R)$ if it is not. For simplicity, we will focus on the case of $SL(5,R)$. The gravitational variables will then be a connection $B^{A}_{\ph{A}B\mu}$ valued in the Lie algebra of $SL(5,R)$ and fields $\{H_{AB},n^{A}\}$, where Greek letters denote spacetime coordinate indices. Then, as in the $SL(3,R)$ case, the vanishing of the entire $SL(5,R)$ curvature two-form can permit the existence of coordinate fields $\xi^{I}$ such that:

\begin{align}
g_{\mu\nu}  = H_{AB}D_{\mu}n^{A}D_{\nu}n^{B} \overset{*}{=} h_{IJ}\partial_{\mu}\xi^{I}\partial_{\nu}\xi^{J}
\end{align}
where $h_{AB}n^{A}=0$ and it has been assumed that $H_{AB}n^{A}n^{B}=1$ and that the `internal metric' $h_{IJ}$ has signature $(-,+,+,+)$. We can furthermore identify the co-tetrad/soldering form as:

\begin{align}
e^{A}_{\mu} &=  ({\cal Y}^{-1})^{A}_{\ph{A}B}D_{\mu}n^{B}
\end{align}
where $
h_{AB} =  \delta_{CD}({\cal Y}^{-1})^{C}_{\ph{C}A}({\cal Y}^{-1})^{D}_{\ph{D}B}$, with $({\cal Y}^{-1})^{A}_{\ph{A}B}n^{B}=0$ and $e^{A}_{\mu}n_{A}=0$, where indices have been assumed to be lowered with $H_{AB}$. Then we may define an $SL(5,R)$ tensor:

\begin{align}
{\cal Q}_{ABC} &= P^{D}_{\ph{D}A}P^{E}_{\ph{D}B}e^{\mu}_{C}Q_{DE\mu}
\end{align}
where $e^{\mu}_{A}e_{\mu}^{C} = P^{A}_{\ph{A}C}$ with

\begin{align}
P^{A}_{\ph{A}B} &= \delta^{A}_{\ph{A}B} -\frac{n^{A}n_{B}}{H_{CD}n^{C}n^{D}}
\end{align}
Then, an example Cartan-geometric action principle for symmetric teleparallelism would be based on the gauge symmetry $SL(5,R)$, with action:

\begin{align}
    S[H_{AB},n^{A},w^{A}_{\ph{A}B},\lambda^{A}_{\ph{A}B},\lambda] = -\frac{1}{16\pi G}&\int d^{4}x\sqrt{-g}\bigg(\frac{1}{4}{\cal Q}_{ABC}{\cal Q}^{ABC}-\frac{1}{2}{\cal Q}_{ABC}{\cal Q}^{CBA}\nn\\
    &\quad\quad\quad\quad\,\,\,\,+\frac{1}{2}{\cal Q}_{A\ph{A}B}^{\ph{A}A}{\cal Q}_{C\ph{C}}^{\ph{C}CB}-\frac{1}{4}{\cal Q}_{AB}^{\ph{AB}A}{\cal Q}_{C\ph{C}B}^{\ph{C}C}\bigg)\nn\\
    +&\int\lambda^{A}_{\ph{A}B}\wedge F^{B}_{\ph{B}A} + \lambda(H_{AB}n^{A}n^{B}-1)
    \label{symtelact}
\end{align}
where indices have been raised with $H^{AB}$, and the Lagrange multiplier two-form $\lambda^{A}_{\ph{A}B}$ enforces the constraint $F^{B}_{\ph{B}A}=0$. This specific combination of quadratic invariants formed from non-metricity is indeed the combination producing the symmetric-teleparallel equivalent of General Relativity \cite{BeltranJimenez:2019esp}. 

Note that as $F^{IJ}_{\ph{IJ}ab}=0$ then from (\ref{hijholo}) the shape change - as measured by $\Delta h_{IJ}$ when the shape is transported around a small loop - will be zero. 
One may ask whether there is an alternative prescription where geometrical information about $M$ is, rather, encoded in holonomies of shape. It can be argued that there are obstructions to this. In this setup the transport/rolling connection takes values in the Cartan decomposition of the Lie algebra
$\mathfrak{sl}(d,\mathbb{R})=\mathfrak{so}(d)\oplus\mathfrak{\sigma}$, where $\mathfrak{\sigma}$
denotes the symmetric traceless matrices generating shears. The number of generators of $\mathfrak{so}(d)$ is $d(d-1)/2$ whilst the number of shear generators is $(d-1)(d+2)/2$, differing by $d-1$, so there is a general mismatch in counting for $d\geq 2$, which reflects the lack of a canonical way to identify elements of $\mathfrak{so}(d)$ with those of $\sigma$. A further obstruction is that 
$\mathfrak{\sigma}$ is not a subalgebra: as may be checked, $[\mathfrak{\sigma},\mathfrak{\sigma}]=\mathfrak{so}(d)$, so
shape holonomy around finite loops cannot be consistently restricted to the shear
sector: the composition of loops regenerates rotations.
 Nonetheless, transport involving shear can contain geometrical information beyond the metric tensor of $M$. In Section \ref{subsection-geometrodynamics}, it will be shown that extrinsic curvature of spacelike hypersurfaces in gravitational theory can be encoded in the change of the spatial metric after parallel transport with a non-metric connection around a small loop.

\begin{figure}[h!]
    \centering
\includegraphics[width=0.65\linewidth]{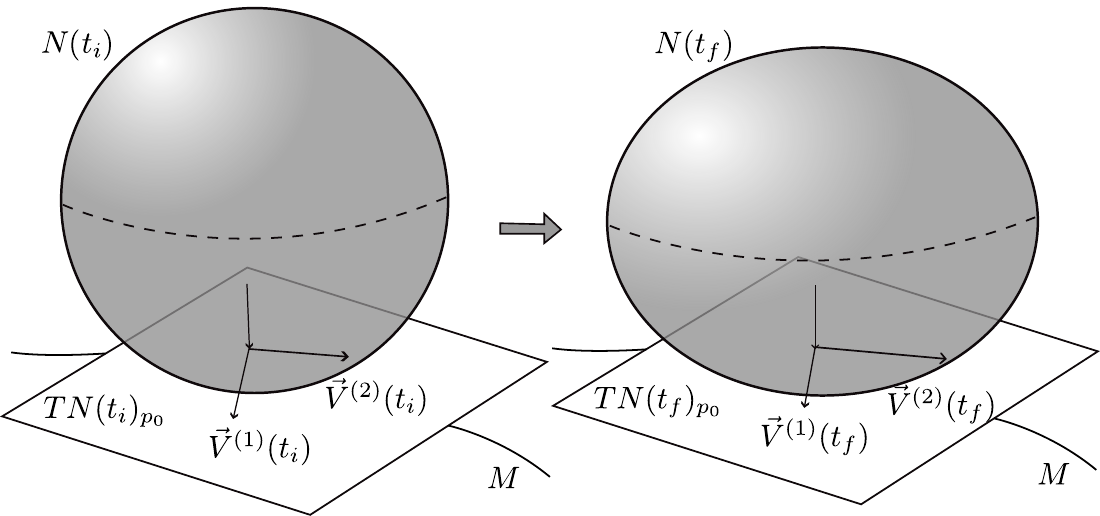}
    \caption{Shape evolution of $N$ after transport around a closed loop on $M$ in the interval $[t_{i},t_{f}]$. Vectors $\vec{V}^{(1)}$ and $\vec{V}^{(2)}$, after transport around the loop, may experience - depending on the form of the transport/rolling connection - no rotation in the embedding space and instead change their size by an amount depending on orientation in the tangent space $TN$.}
    \label{fig:nonmetricity}
\end{figure}

\subsection{Gravity as a gauge theory of the Poincar\'{e} group}
\label{subsection-poincare}
The MacDowell-Mansouri approach to gravity presents gravity as a spontaneously broken gauge theory of groups $SO(1,4)$ or $SO(2,3)$. Another popular approach is to regard the gauge symmetry as instead being that of the Poincar\'{e} group $ISO(1,3)$ \cite{Kibble:1961ba}. We now illustrate how Poincar\'{e} gauge theory can be regarded as a spontaneously broken gauge theory of the group $SL(5,R)$ with some constraints placed on the curvature. As in the previous subsection, the dynamical fields will be a symmetric $SL(5,R)$ tensor $H_{AB}$, an $SL(5,R)$ vector $n^{A}$, and a connection $B^{A}_{\ph{A}B\mu}\in \mathfrak{sl}(5,R)$. We may decompose $B^{A}_{\ph{A}B\mu}$ as follows:
\begin{align}
B^{A}_{\ph{A}Ba} &\overset{*}{=}  \begin{pmatrix}
-\frac{1}{4}c_{a}\delta^{I}_{\ph{I}J} + w^{I}_{\ph{I}Ja } & E^{I}_{a} \\
V_{Ia} & c_{a}
\end{pmatrix}  \label{bcon3}
\end{align}
where $I,J,K,\dots= 0,\dots, 3$ and $w^{I}_{\ph{I}J\mu}\in \mathfrak{sl}(4,R)$. The field $w^{I}_{\ph{I}J\mu}$ can further be decomposed into symmetric $S^{IJ}_{\ph{IJ}\mu}$ and anti-symmetric parts $A^{IJ}_{\ph{IJ}\mu}$ (with index assumed raised with the inverse matrix of $H_{IJ}$). Given this decomposition of $B^{A}_{\ph{A}B\mu}$ the curvature can be decomposed as in equations (\ref{fij})-(\ref{curvature4}), with adjustment made for the differing coefficient in front of the part of the connection proportional to $\delta^{I}_{\ph{I}J}$. If we assume that $n^{A}$ is non-vanishing and that we can identify a gauge in which $n^{A} \propto  \delta^{A}_{\ph{A}4}$ then if the following conditions hold:

\begin{align}
F^{4}_{\ph{4}Jab} &=0 \label{pcurvcon1}\\
F^{4}_{\ph{4}4ab} &= 0 \label{pcurvcon2}\\
F^{IJ}_{\ph{IJ}ab} &= - F^{JI}_{\ph{JI}ab}  \label{pcurvcon3}
\end{align}
then $\{c_{\mu},V_{J\mu},S^{I}_{\ph{I}J\mu}\}$ become `pure gauge' and we can locally look to further gauge fix so that $B^{A}_{\ph{A}B\mu}$ achieves the following form:

\begin{align}
B^{A}_{\ph{A}B\mu} &\overset{*}{=}  \begin{pmatrix}
  A^{I}_{\ph{I}J\mu } & E^{I}_{\mu} \\
0 & 0
\end{pmatrix}   \in \mathfrak{iso}(1,3)
\end{align}
This form of the connection is in the Lie algebra of $ISO(1,3)$ and it is also a non-metric connection with respect to $H_{AB}$ in that - assuming the possibility of a further gauge fixing 

\begin{align}
H_{AB}\overset{*}{=} \ell^{2}\mathrm{diag}(-1,1,1,1,1)
\label{poinHgauge}
\end{align}
where $\ell$ is a constant. It follows so that $H_{IJ}= \ell^{2}\eta_{IJ}$ and where $\eta_{IJ}$ is the invariant matrix of $SO(1,3)$. It follows then that:

\begin{align}
Q_{4I\mu} &\overset{*}{=} \ell^{2}E_{I\mu} \label{pattern}
\end{align}
This suggests an interpretation of the Poincar\'{e} soldering form $E^{I}_{\mu}$ as non-metricity with respect to the extended group. Related relations between general-linear and Lorentz-Poincar{\'e} geometries have been discussed in Refs.~\cite{Koivisto:2019ejt,Wheeler:2024iql}. Here the distinguished direction $n^A$ and the curvature constraints locally reduce the $SL(5,R)$ connection to its $ISO(1,3)$ sector.

This construction can be compared with different patterns isolating the Lorentz--Poincar\'e geometry from a general-linear one. The
general-linear Cartan-khronon construction uses a symmetry-breaking field to select the Lorentz subgroup of $GL(4,R)$, thereby splitting the $GL$ connection into a $SO$ connection and components along the
broken generators \cite{Koivisto:2019ejt}. 
At the level of the metric-affine structure
equations, a closely related correspondence was exhibited in Ref.~\cite{Wheeler:2024iql}: the mixed-symmetry part of non-metricity
can be absorbed into a redefinition of torsion, leaving the structure
equations of a metric-compatible Poincar\'e geometry.\footnote{The
correspondence has also been extended to the spinorial sector
\cite{Wheeler:2026xfx}. A fully dynamical realization along the
Cartan-khronon lines would additionally require the corresponding
symmetry-breaking sector and Wick prescription.}
The present $SL(5,R)$ construction realizes the same basic
reduction mechanism with a different enlarged algebra. The distinguished
direction $n^{A}$ induces a $4+1$ decomposition of the connection, while
the curvature constraints remove the complementary fields
$\{c_{\mu},V_{I\mu},S_{IJ\mu}\}$ locally and leave the
$ISO(1,3)$ sector. In this decomposition the off-diagonal component
$E^{I}_{\mu}$ becomes the Poincar\'e translational gauge field, or
soldering form, and at the same time appears as an extended
non-metricity component through (\ref{pattern}).

A candidate action for General Relativity is then:

\begin{align}
S[n,H,B,\lambda,\xi] &= \int \eps_{ABCDE}H^{DF}n^{E}Dn^{A}\wedge Dn^{B} \wedge F^{C}_{\ph{C}F}  \nn\\
&+  \lambda (H_{AB}n^{A}n^{B}-\ell^{2}) 
+\zeta (\det{H}+\ell^{10})+\xi^{B} F^{A}_{\ph{A}B}n_{A} +\xi^{A}_{\ph{A}B}F^{B}_{\ph{B}A}
\end{align}
where $H^{AB}$ is the matrix inverse of $H_{AB}$, $n_{A}\equiv H_{AB}n^{B}$,  $\xi^{AB}\equiv \xi^{A}_{\ph{A}C}H^{CB} = \xi^{BA}$. The Lagrangian constraint imposed via the $\lambda$ Lagrange multiplier represents a fixed norm constraint on $n^{A}$ with respect to the object $H_{AB}$, the Lagrangian constraint imposed via the $\zeta$ Lagrange multiplier restricts $H_{AB}$ to be unimodular, justifying the above gauge fixing, whilst the equations of motion resulting from stationarity of the action under variations of $\xi^{A}$ and $\xi^{A}_{\ph{A}B}$ collectively covariantly implement the curvature constraints (\ref{pcurvcon1}-\ref{pcurvcon3}). 

The constraint enforced by $\lambda$ may alternatively be implemented at the level of the action. Given (\ref{poinHgauge}), a further partial gauge fixing is

\begin{align}
n^{A} &\overset{*}{=} \delta^{A}_{\ph{A}4} \label{poinngauge}
\end{align}
Note that in this gauge we have $Dn^{I} \overset{*}{=} E^{I}$. We may employ the gauge fixing conditions (\ref{poinHgauge}) and (\ref{poinngauge}) at the level of the action\footnote{In this case gauge fixing at the level of the action and then determining the field equations is equivalent to determining the full field equations and then gauge fixing at the level of the equations of motion \cite{Motohashi:2016prk}. This is because the gauge functions required to implement (\ref{poinHgauge}) and (\ref{poinngauge}) are fully determined by these gauge conditions.}, and recover the action:

\begin{align}
S[A,E,c,V,S,\xi] &= \frac{1}{\ell^{2}}\int \eps_{IJKL} E^{I}\wedge E^{J} \wedge F^{[KL]} +\xi^{I} F^{4}_{\ph{4}I} +(\xi_{4}+\xi_{44})F^{44} + \xi_{IJ}F^{(IJ)}
\end{align}
In this formulation, following gauge fixing, we may identify $E^{I}$ with the co-tetrad and the field $A^{I}_{\ph{I}J}$ with the spin connection of Einstein-Cartan gravity. The additional constraints on the curvature essentially allow, at the level of the equations of motion, the remaining parts $\{c,V^{I},S^{IJ}\}$ of the $SL(5,R)$ connection to be locally set to zero.

\subsection{Shape holonomy and geometrodynamics}
\label{subsection-geometrodynamics}

We now show that the Einstein-Cartan formulation of General Relativity in `3+1' form can be cast in a Cartan-geometric form with a non-metrical connection with non-vanishing curvature two-form. 

Consider a foliation of spacetime into a  time function $t$ and spatial hypersurfaces $\Sigma_{t}$ with coordinates $x^{a}$. We will look to describe geometry in terms of dynamical fields $\{n^{A},H_{AB}\}$ taken to transform respectively as a vector and tensor under $GL(4,R)$ gauge transformations. The final dynamical field will be a connection $B^{A}_{\ph{A}B\mu}$ valued in the Lie algebra of $GL(4,R)$. We will look to identify the \emph{spacetime} metric with:

\begin{align}
g_{\mu\nu} &= H_{AB} D_{\mu}n^{A}D_{\nu}n^{B} \label{glmetdef}
\end{align}
where $H_{AB}$ is the analogue of (\ref{HABdef}) with $\alpha$ assumed to be non-zero. We will assume that the field $n^{A}$ has fixed norm $H_{AB}n^{A}n^{B}=-\ell^{2}$ - where $\ell$ is a constant - and that $\det[H]<0$ (with one negative eigenvalue) and so one can locally find a $GL(4,R)$ gauge where:

\begin{align}
H_{AB}  &\overset{*}{=} \ell^{2}\mathrm{diag}(-1,1,1,1) \equiv \ell^{2}\eta_{AB} \label{gauge_fix_h}\\
n^{A} &\overset{*}{=}  \delta^{A}_{\ph{A}0} \label{gauge_fix_n}
\end{align}
From (\ref{glmetdef}), we have then that

\begin{align}
g_{\mu\nu} &\overset{*}{=} \ell^{2}\eta_{AB}B^{A}_{\ph{A}0\mu}B^{B}_{\ph{B}0\nu}
\end{align}
This suggests that in this gauge we can identify $\ell B^{A}_{\ph{A}0\mu}$  with the spacetime co-tetrad/soldering form $e^{A}_{\mu}$ and that

\begin{align}
B^{A}_{\ph{A}B\mu} &\overset{*}{=} \begin{pmatrix}
\frac{1}{\ell}e^{0}_{\mu} & K_{I\mu} \\
\frac{1}{\ell}e^{I}_{\mu} & \Gamma^{I}_{\ph{I}J\mu} 
\end{pmatrix} \label{buh}
\end{align}
where $I,J,K,\dots = 1,2,3$ and $K_{I\mu} = B^{0}_{\ph{0}I\mu}$. We may define an additional object $\omega^{A}_{\ph{A}B\mu}$ that also transforms as a $GL(4,R)$ connection:

\begin{align}
\omega^{A}_{\ph{A}B} & \equiv B^{A}_{\ph{A}B}  -L^{A}_{\ph{A}B}
\end{align}
where 
\begin{align}
L^{A}_{\ph{A}B} &=  \ell^{-4}\big(\ell^{2}H^{AE}+ \frac{1}{2}n^{A}n^{E}\big)n^{C}n_{B}DH_{EC}
\end{align}
and where $n_{A}\equiv H_{AB}n^{B}$. In the gauge defined by (\ref{gauge_fix_h}) and (\ref{gauge_fix_n}) we have that:

\begin{align}
\omega^{AB}_{\ph{AB}\mu} & \overset{*}{=}  \begin{pmatrix}
0 & K^{I}_{\mu} \\
-K^{I}_{\mu} & \Gamma^{IJ}_{\ph{IJ}\mu}
\end{pmatrix} 
\end{align}
where $\omega^{AB}_{\ph{AB}\mu} = H^{BC}\omega^{A}_{\ph{A}C\mu}$. If, furthermore, $\Gamma^{(IJ)}_{\ph{(IJ)}\mu}=0$ then $\omega^{AB}_{\ph{AB}\mu}$ is valued in the Lie algebra of $SO(1,3)$ as is the case for the spin connection in the Einstein-Cartan formulation of General Relativity. We now look to construct an action for the gravitational fields so that the equations of motion are those of the Einstein-Cartan formulation of General Relativity. Consider the following action:

\begin{align}
S[n,H,B,\lambda] &= \int \sqrt{-H}\eps_{ABCD}H^{DE}Dn^{A}\wedge Dn^{B} \wedge {\cal F}^{C}_{\ph{C}E}(\omega) + \lambda (H_{AB}n^{A}n^{B}+\ell^{2}) \label{Hact}
\end{align}
where

\begin{align}
{\cal F}^{A}_{\ph{A}B} &= d\omega^{A}_{\ph{A}B}+\omega^{A}_{\ph{A}C}\wedge \omega^{C}_{\ph{C}B} \\
&= F^{A}_{\ph{A}B}(B) - D^{(B)}L^{A}_{\ph{A}C} + L^{A}_{\ph{A}C}\wedge L^{C}_{\ph{C}B}
\end{align}
The Lagrangian constraint associated with the field $\lambda$ fixes $H_{AB}n^{A}n^{B}=-\ell^{2}$. We may employ the gauge fixing conditions (\ref{gauge_fix_h}) and (\ref{gauge_fix_n}) at the level of the action \footnote{As in Subsection \ref{subsection-poincare}, gauge fixing at the level of the action and then determining the field equations is equivalent to determining the full field equations and then gauge fixing at the level of the equations of motion \cite{Motohashi:2016prk}. This is because the gauge functions required to implement (\ref{gauge_fix_h}) and (\ref{gauge_fix_n}) are fully determined by these gauge conditions.}, and recover the action:

\begin{align}
S[e,K,\Gamma] &\overset{*}{\propto} \int \eps_{ABCD}\eta^{DE}e^{A}\wedge e^{B} \wedge {\cal F}^{C}_{\ph{C}E}(\omega) 
\end{align}
where recall that $e^{A}_{\mu} \overset{*}{=} B^{A}_{\ph{A}0\mu}$, $\omega^{0}_{\ph{0}I\mu}\overset{*}{=}B^{0}_{\ph{0}I\mu}$ and $\omega^{I}_{\ph{I}J\mu} \overset{*}{=} B^{I}_{\ph{I}J\mu}$. Hence, we recover the Einstein-Cartan action so for example varying with respect to $\omega^{A}_{\ph{A}B\mu}$ we recover the equation of motion:

\begin{align}
de^{A} + \omega^{A}_{\ph{A}B}\wedge  e^{B} &=0
\end{align}
Solving this implies that $\omega^{A}_{\ph{A}B\mu}$ takes the form of the torsion-free spin connection, with the pullback $K_{Ia}$ to surfaces of constant time encoding the extrinsic curvature of these surfaces in spacetime and with the pullback $\Gamma^{IJ}_{\ph{IJ}a}=-\Gamma^{JI}_{\ph{JI}a}$ representing the  torsion-free and metric-compatible spin connection. 

An interesting interpretation is then possible in terms of the original connection $B^{A}_{\ph{A}B\mu}$. From (\ref{fij}) and (\ref{hijholo})  we have that the parallel transport of $h_{AB}$ via the connection $B^{A}_{\ph{A}B\mu}$ around a small loop taking the form of a parallelogram defined by infinitesimal vectors $\vec{\delta x} = \delta x^{a}\partial_{a}$ and $\vec{\delta y} = \delta y^{a} \partial_{a}$ on $\Sigma_{t}$ yields a change in $H^{IJ}$ equal to:

\begin{align}
\Delta H^{IJ} &=  \frac{4}{\ell}e^{(I}_{\ph{(I}[a|}K^{J)}_{\ph{J)}|b]}\delta x^{a}\delta y^{b}  \label{DeltaHIJ}
\end{align}
where $K^{I}_{a}\equiv H^{IJ}K_{Ja}$. Therefore the change in $H_{IJ}$, which can be interpreted as a change in shape of a three dimensional ellipsoid when rolled on $\Sigma_{t}$ (both embedded in a flat embedding space) around a small loop, is related to the extrinsic curvature $K_{ab}$ of $\Sigma_{t}$ in spacetime. Interestingly, the change $\Delta H^{IJ}$ is not sensitive to all of $K^{I}_{\ph{I}a}$. Specifically, if $K^{I}_{\ph{I}a}$ is decomposed as follows:

\begin{align}
K^{I}_{\ph{I}a} = \frac{K}{3}e^{I}_{a}  + \hat{K}^{I}_{\ph{I}a}
\end{align}
where $K\equiv e^{a}_{I}K^{I}_{\ph{I}a}$ and $e^{a}_{I}\hat{K}^{I}_{\ph{I}a}=0$ then only $\hat{K}^{I}_{\ph{I}a}$ contributes to (\ref{DeltaHIJ}) as $e^{(I}_{[a}e^{J)}_{b]}=0$.

\subsubsection{Analogous construction in Einstein-Cartan gravity}

The same structure can be recovered from within the context of Einstein-Cartan gravity based on fields $\{e^{A}_{\mu},\omega^{AB}_{\ph{AB}\mu}\}$. In this case, a field $N^{A}$ which is constrained to be unit timelike with respect to the $SO(1,3)$-invariant metric $\eta_{AB}$ is known to be useful in defining the `$3+1$' decomposition of $\{e^{A}_{\mu},\omega^{AB}_{\ph{AB}\mu}\}$ \cite{Peldan:1993hi}:

\begin{align}
e^{A}_{a} &=  V^{A}_{a}\\
e^{A}_{t} &= N N^{A} + N^{a}V^{A}_{a}
\end{align}
where coordinates $x^{a}$ denote spatial coordinates on surfaces of constant coordinate time $t$ and
where $V^{A}_{a}N_{A}=0$. Then the metric tensor is:

\begin{align}
g_{\mu\nu} &= \begin{pmatrix}
-N^{2}+N_{a}N^{a} & N_{a}\\
N_{a} & V^{A}_{a}V_{Ab}
\end{pmatrix}
\end{align}
where $N_{a} = V_{Aa}V^{A}_{b}N^{b}$, which is recognizable as the ADM form of the metric tensor with the identification of $V^{A}_{a}V_{Ab}=q_{ab}$ as the metric on surfaces of constant coordinate time $t$. The field $N^{A}$ can further be used to decompose the spin connection $\omega^{AB}_{\ph{AB}\mu}$ into a `boost' part $K^{A}_{\mu}$ and a rotation part $\Gamma^{AB}_{\ph{AB}\mu}$ where $K^{A}_{\mu}N_{A}=\Gamma^{AB}_{\ph{AB}\mu}N_{B}=0$:

\begin{align}
    \omega^{AB}_{\ph{AB}\mu} &= \Gamma^{AB}_{\ph{AB}\mu} + 2N^{[A}K^{B]}_{\mu}\\
    &\overset{*}{=} \begin{pmatrix}
    0 & K^{I}_{\mu} \\
    -K^{I}_{\mu} & \Gamma^{IJ}_{\ph{IJ}\mu}
    \end{pmatrix} \label{mateq1}
\end{align}
where equation (\ref{mateq1}) gives the form of $\omega^{AB}_{\ph{AB}\mu}$ in the gauge $N^{A} \overset{*}{=} \delta^{A}_{\ph{A}0}$ and where `$SO(3)$' indices $1,2,3$ have been labelled $I,J,K$.  Then, the following object may be defined:

\begin{align}
{\cal B}^{AB}_{\ph{AB}\mu} &= \omega^{AB}_{\ph{AB}\mu} + \left(-\frac{1}{\ell}e^{A}_{\mu} + D^{(\omega)}_{\mu}N^{A}\right)N^{B}\label{bconredef}
\end{align}
where $D^{(\omega)}N^{A}\equiv dN^{A}+\omega^{A}_{\ph{A}B}N^{B}$ and $\ell$ is a constant.
Again adopting the  gauge fixing $N^{A} = \delta^{A}_{\ph{A}0}$ we have:

\begin{align}
{\cal B}^{AB}_{\ph{AB}\mu} 
&\overset{*}{=} \begin{pmatrix}
    -\frac{1}{\ell}e^{0}_{\mu} &K^{I}_{\mu} \\
   -\frac{1}{\ell} e^{I}_{\mu} & \Gamma^{IJ}_{\ph{IJ}\mu}
    \end{pmatrix} \label{mateq2}
\end{align}
This illustrates that the gravitational fields $\{e^{A}_{\mu},\omega^{AB}_{\ph{AB}\mu}\}$ can be `packaged together' into a single object and that in this gauge the object corresponds to (\ref{buh}) which was associated with a $GL(4,R)$ connection.
The field ${\cal B}^{AB}_{\ph{AB}\mu}$ - due to it differing from $\omega^{AB}_{\ph{AB}\mu}$ by Lorentz tensor-valued one-forms via (\ref{bconredef}) - transforms inhomogeneously as a connection for the group $SO(1,3)$ whilst possessing non-metricity:

\begin{align}
Q_{AB\mu} &\equiv -D_{\mu}^{({\cal B})}\eta_{AB} =  \begin{pmatrix}
\frac{2}{\ell}e_{0\mu} & \frac{1}{\ell}e_{I\mu} -K_{I\mu}\\
\frac{1}{\ell}e_{I\mu}-K_{I\mu} &0
\end{pmatrix}
\end{align}
where indices have been lowered with $\eta_{AB}$. 

Therefore in this gauge the form $Q_{00}$ is related to $e_{0}$ (and hence the lapse function) whilst the pullback of $Q_{OI}$ to surfaces of constant time is proportional to the difference between the extrinsic curvature of spacetime and the extrinsic curvature of de Sitter  space of radius $\ell$ in flat slicing. Note that although the components $Q_{IJ}$ are zero, the change in $\eta^{IJ}=\delta^{IJ}$ when parallel transported according to ${\cal B}^{A}_{\ph{A}B\mu}$ around the small loop defined by infinitesimal vectors $\vec{\delta x} = \delta x^{a}\partial_{a}$ and $\vec{\delta y} = \delta y^{a} \partial_{a}$ is:

\begin{align}
\Delta^{({\cal B})}\eta^{IJ} &=  \frac{4}{\ell}e^{(I}_{\ph{(I}[a|}K^{J)}_{\ph{J)}|b]}\delta x^{a}\delta y^{b}  \label{DeltaetaIJ}
\end{align}
where here $K^{I}_{a}\equiv \eta^{IJ}K_{Ja}$.
This result is analogous to that of (\ref{DeltaHIJ}), and expresses the change of `shape' of the spatial part of the internal metric in terms of the scale $\ell$, the spatial frame field $e^{I}_{a}$ and the trace-free part of the extrinsic curvature $K^{I}_{a}$.

Formally, for transport around finite paths on surfaces of constant $t$,  the result for the finite change -i.e. holonomy - $\Delta_{F}^{({\cal B})} \eta^{IJ}$ is expressible in terms of a `surface' ordered exponential of an integration of the curvature two-form over the finite loop. This will generally produce contributions to the holonomy of other parts of the curvature of ${\cal B}$ than are present in the infinitesimal case, obscuring the relation between the change of $\delta^{IJ}$ (or in the $GL(4,R)$ case $H^{IJ}$) and extrinsic curvature.

\section{Discussion and Conclusions}
\label{section-discussionandconclusions}

Our results illustrate that the geometrical trinity of gravity admits a Cartan geometric interpretation as a set of gauge theories with constraints present on the Higgs sector of the theory and - as in the case of teleparallelism, symmetric teleparallelism and gravity regarded as a gauge theory of the Poincar\'{e} group -  constraints on the curvature of the gravitational gauge field. 

These results underline the remarkable array of formulations that General Relativity admits. A potential utility of these different formulations is that they may provide different starting points for exploring extensions to General Relativity. Whilst writing this article, an independent work \cite{Capozziello:2026pys} focusing on - and also confirming the existence of - a Cartan-geometrical description of the geometrical trinity of gravity has become known to the authors\footnote{The authors of \cite{Capozziello:2026pys} describe their approach as being \emph{pre-geometrical} in the sense that the metric tensor is defined from more basic objects, which, using the nomenclature of the present manuscript, are fields $\{H_{AB},n^{A},B^{A}_{\ph{A}B\mu}\}$.}; this work emphasizes that not just General Relativity but also modified theories of gravity can be recovered via torsion and non-metricity appearing in different combinations and in different functional forms from those in actions such as (\ref{telact}) and (\ref{symtelact}).

Another step in the direction of generalization is to reduce the number of Lagrangian constraints appearing in actions such as (\ref{actsw}), (\ref{telact}), and (\ref{symtelact}). For example, it is known that allowing the field $n^{A}$ in the Stelle-West model of gravity to be unconstrained allows for the construction of polynomial actions where the now `unfrozen' scalar degree of freedom present in $n^{A}$ produces a scalar field with canonical kinetic term and Peebles-Ratra type potential \cite{Westman:2013mf}. It is then natural to wonder, for example, about the dynamics of $n^{A}$ in Cartan-geometrical formulations of teleparallel and symmetric-teleparallel gravity if it were also allowed to freely vary. This is left as a topic for further exploration.

In addition to the previous results regarding the geometrical trinity of gravity, in Section \ref{subsection-geometrodynamics} it was shown that the Cartan-geometrical approach suggested a connection redefinition (\ref{bconredef}) within the framework of Einstein-Cartan gravity, where the connection ${\cal B}^{AB}$ possesses non-metricity and, in the context of the $3+1$ decomposition of spacetime, this results in non-preservation of the internal $SO(3)$ metric $\eta_{IJ}$ under parallel transport around an infinitesimal loop with the degree of non-preservation being related to the trace-free part of the extrinsic curvature of spacelike hypersurfaces. In this sense, the trace-free part of the extrinsic curvature of such surfaces embedded in spacetime may be interpreted as the shape evolution of a surface parallel transported - via a Cartan connection - on that same hypersurface if embedded in a higher dimensional ambient flat spacetime. 

Motivated by the interpretation of non-metricity in terms of transport of a higher dimensional space on a lower dimensional manifold, it would be interesting to explore the geometrical interpretation of systems where a connection with non-metricity is known to arise. Of particular interest is the occurrence of structure similar to co-tetrads and torsion in condensed matter systems \cite{Volovik:2023faj} and the recent discovery  that non-metric connections naturally arises in the description of effective acoustic spacetimes in scalar-tensor theories \cite{Sawicki:2024ryt}.

By way of generalization, one could look in the embedding picture to extend results to the case where shape evolution depends not just globally on the position of the shape its path on $M$, but also on the position of points on the rolling manifold $N$. This would involve the rolling of more general shapes than ellipsoids and it would be interesting to see whether new shape degrees of freedom possess an intrepretation in terms of quantities on $M$.

\paragraph{Acknowledgements} We thank Paola Delgado for motivating this research by raising the question of how to interpret non-metricity and symmetric teleparallelism in a Cartan-geometrical framework. We also thank Amel Durakovic, Priidik Gallagher, and Hans Westman for helpful discussions.

\appendix
\section{The orbit of the group $SL(m,R)$}
\label{A1}

A general element $M^{A}_{\ph{A}B}$ of the group $SL(m,R)$ can be written using the Iwasawa decomposition:

\begin{align}
M &= KAN \label{iwasawa_decomposition}
\end{align}
where $K$ are orthogonal matrices, $A$ are diagonal matrices with positive entries and unit determinant, and $N$ belongs to the unipotent group. For example, for $SL(2,R)$ we have that:

\begin{align}
K &= \begin{pmatrix}
\cos(\phi) & -\sin(\phi) \\
\sin(\phi)  & \cos(\phi)
\end{pmatrix}  \quad A = \begin{pmatrix}
\alpha & 0\\
0  & \alpha^{-1}
\end{pmatrix}\quad N= \begin{pmatrix}
1 & N^{1}_{\ph{1}2}\\
0 & 1
\end{pmatrix} 
\end{align}
where $\phi,\alpha,N^{1}_{\ph{1}2}$ are real numbers and $\alpha > 0$.

We are interested in determining the orbit of a vector $V^{A} = \delta^{A}_{\ph{A}m} = (0,\dots,0,1)$ i.e. the space of values that the vector $MV$ can take. Using the decomposition (\ref{iwasawa_decomposition}) we can first evaluate:

\begin{align}
N^{A}_{\ph{A}B}V^{B} &= (N^{1}_{\ph{1}m},N^{2}_{\ph{2}m},\dots, 1)
\end{align}
Then

\begin{align}
A^{A}_{\ph{A}B}N^{B}_{\ph{B}C}V^{C} &= (\lambda_{1}N^{1}_{\ph{1}m},\lambda_{2}N^{2}_{\ph{2}m},\dots, \lambda_{m})
\end{align}
where due to the unit-determinant condition we have that $\lambda_{m}=1/\prod_{i=1}^{m-1}\lambda_{i}$. Given that $\lambda_{A} > 0$, and $N^{i<m}_{\ph{i<m}m}$ are arbitrary, we can choose all components of $ANV$ to be greater than or equal to zero and so identify the vector $ANV$ with a point in $\mathbb{R}^{m}$ coordinatized by $x^{A}$ where all $x^{A}\geq 0$ and $\delta_{AB}x^{A}x^{B}>0$ due to the condition $\lambda_{m}\neq 0$. All remaining points in $\mathbb{R}^{m}$ (with the exception of the origin) can then be reached by a rotation represented by $K\in SO(m)$ hence we identify the orbit of $V$ with $\mathbb{R}^{m}\setminus\{0\}$.

\bibliographystyle{unsrt}
\bibliography{references}

@article{Koivisto:2026rmp,
    author = "Koivisto, Tomi S.",
    title = "{Hamiltonian surface charges in general parallel relativity}",
    doi = "10.1007/s10714-026-03602-6",
    journal = "Gen. Rel. Grav.",
    volume = "58",
    number = "9",
    pages = "103",
    year = "2026"
}

@article{Wise:2006sm,
    author = "Wise, Derek K.",
    title = "{MacDowell-Mansouri gravity and Cartan geometry}",
    eprint = "gr-qc/0611154",
    archivePrefix = "arXiv",
    doi = "10.1088/0264-9381/27/15/155010",
    journal = "Class. Quant. Grav.",
    volume = "27",
    pages = "155010",
    year = "2010"
}

@article{Wheeler:2026xfx,
    author = "Wheeler, James T.",
    title = "{Dirac sources for nonmetricity and torsion in metric-affine gravity}",
    eprint = "2601.09013",
    archivePrefix = "arXiv",
    primaryClass = "gr-qc",
    doi = "10.1140/epjc/s10052-026-15696-y",
    journal = "Eur. Phys. J. C",
    volume = "86",
    number = "5",
    pages = "484",
    year = "2026"
}

@article{Barker:2022jsh,
    author = "Barker, W. E. V.",
    title = "{Geometric multipliers and partial teleparallelism in Poincar{\'e} gauge theory}",
    eprint = "2205.13534",
    archivePrefix = "arXiv",
    primaryClass = "gr-qc",
    doi = "10.1103/PhysRevD.108.024053",
    journal = "Phys. Rev. D",
    volume = "108",
    number = "2",
    pages = "024053",
    year = "2023"
}

@article{Wheeler:2024iql,
    author = "Wheeler, James T.",
    title = "{Poincare gauge gravity from nonmetric gravity}",
    eprint = "2407.13867",
    archivePrefix = "arXiv",
    primaryClass = "gr-qc",
    doi = "10.1016/j.nuclphysb.2025.116860",
    journal = "Nucl. Phys. B",
    volume = "1014",
    pages = "116860",
    year = "2025"
}

@article{Koivisto:2019ejt,
    author = "Koivisto, Tomi and Hohmann, Manuel and Z{\l}o{\'s}nik, Tom",
    title = "{The General Linear Cartan Khronon}",
    eprint = "1905.02967",
    archivePrefix = "arXiv",
    primaryClass = "gr-qc",
    reportNumber = "NORDITA 2019-045",
    doi = "10.3390/universe5070168",
    journal = "Universe",
    volume = "5",
    number = "6",
    pages = "168",
    year = "2019"
}

@article{BeltranJimenez:2020sih,
    author = "Beltr{\'a}n Jim{\'e}nez, Jose and Heisenberg, Lavinia and Koivisto, Tomi",
    title = "{The coupling of matter and spacetime geometry}",
    eprint = "2004.04606",
    archivePrefix = "arXiv",
    primaryClass = "hep-th",
    doi = "10.1088/1361-6382/aba31b",
    journal = "Class. Quant. Grav.",
    volume = "37",
    number = "19",
    pages = "195013",
    year = "2020"
}

@article{Lindwasser:2022nfa,
    author = "Lindwasser, Lukas W. and Tomboulis, E. T.",
    title = "{Searching for gravity without a metric}",
    eprint = "2207.01067",
    archivePrefix = "arXiv",
    primaryClass = "hep-th",
    doi = "10.1103/PhysRevD.106.084026",
    journal = "Phys. Rev. D",
    volume = "106",
    number = "8",
    pages = "084026",
    year = "2022"
}

@book{Eisenhart1927,
  author    = {Luther Pfahler Eisenhart},
  title     = {Non-Riemannian Geometry},
  series    = {American Mathematical Society Colloquium Publications},
  volume    = {8},
  publisher = {American Mathematical Society},
  address   = {New York},
  year      = {1927},
  pages      = {vii+184},
}

@phdthesis{Iosifidis:2019dua,
    author = "Iosifidis, Damianos",
    title = "{Metric-Affine Gravity and Cosmology/Aspects of Torsion and non-Metricity in Gravity Theories}",
    eprint = "1902.09643",
    archivePrefix = "arXiv",
      journal = "PhD Thesis",
    primaryClass = "gr-qc",
    year = "2019"
}

@article{Iosifidis:2023eom,
    author = "Iosifidis, Damianos and Hehl, Friedrich W.",
    title = "{Motion of test particles in spacetimes with torsion and nonmetricity}",
    eprint = "2310.15595",
    archivePrefix = "arXiv",
    primaryClass = "gr-qc",
    doi = "10.1016/j.physletb.2024.138498",
    journal = "Phys. Lett. B",
    volume = "850",
    pages = "138498",
    year = "2024"
}

@article{Hehl1976b,
  author  = {Hehl, Friedrich W. and Kerlick, G. David and von der Heyde, Paul},
  title   = {On Hypermomentum in General Relativity. II. The Geometry of Space-Time},
  journal = {Zeitschrift für Naturforschung A},
  volume  = {31},
  number  = {6},
  pages   = {524--527},
  year    = {1976},
  doi     = {10.1515/zna-1976-0602}
}

@article{Hehl1995,
  author  = {Hehl, Friedrich W. and McCrea, J. Dermott and Mielke, Eckehard W. and Ne'eman, Yuval},
  title   = {Metric-affine gauge theory of gravity: Field equations, Noether identities, world spinors, and breaking of dilation invariance},
  journal = {Physics Reports},
  volume  = {258},
  number  = {1--2},
  pages   = {1--171},
  year    = {1995},
  doi     = {10.1016/0370-1573(94)00111-F}
}

@article{Hehl1976a,
  author  = {Hehl, Friedrich W. and Kerlick, G. David and von der Heyde, Paul},
  title   = {On Hypermomentum in General Relativity. I. The Notion of Hypermomentum},
  journal = {Zeitschrift für Naturforschung A},
  volume  = {31},
  number  = {2},
  pages   = {111--114},
  year    = {1976},
  doi     = {10.1515/zna-1976-0201}
}

@article{Percacci:2009ij,
    author = "Percacci, R.",
    editor = "Pinheiro, Carlos and de Arruda, Alberto S. and Blas, Harold and Pires, Gentil O.",
    title = "{Gravity from a Particle Physicists' perspective}",
    eprint = "0910.5167",
    archivePrefix = "arXiv",
    primaryClass = "hep-th",
    reportNumber = "PI-PARTPHYS-156",
    doi = "10.22323/1.081.0011",
    journal = "PoS",
    volume = "ISFTG",
    pages = "011",
    year = "2009"
}

@article{BeltranJimenez:2019esp,
    author = "Beltr{\'a}n Jim{\'e}nez, Jose and Heisenberg, Lavinia and Koivisto, Tomi S.",
    title = "{The Geometrical Trinity of Gravity}",
    eprint = "1903.06830",
    archivePrefix = "arXiv",
    primaryClass = "hep-th",
    doi = "10.3390/universe5070173",
    journal = "Universe",
    volume = "5",
    number = "7",
    pages = "173",
    year = "2019"
}

@article{Leclerc:2005qc,
    author = "Leclerc, M.",
    title = "{The Higgs sector of gravitational gauge theories}",
    eprint = "gr-qc/0502005",
    archivePrefix = "arXiv",
    doi = "10.1016/j.aop.2005.08.009",
    journal = "Annals Phys.",
    volume = "321",
    pages = "708--743",
    year = "2006"
}

@article{Gallagher:2022kvv,
    author = "Gallagher, Priidik and Koivisto, Tomi and Marzola, Luca",
    title = "{Pregeometric first order Yang-Mills theory}",
    eprint = "2202.05657",
    archivePrefix = "arXiv",
    primaryClass = "hep-th",
    doi = "10.1103/PhysRevD.105.125010",
    journal = "Phys. Rev. D",
    volume = "105",
    number = "12",
    pages = "125010",
    year = "2022"
}

@article{Baez:2012bn,
    author = "Baez, John C. and Wise, Derek K.",
    title = "{Teleparallel Gravity as a Higher Gauge Theory}",
    eprint = "1204.4339",
    archivePrefix = "arXiv",
    primaryClass = "gr-qc",
    doi = "10.1007/s00220-014-2178-7",
    journal = "Commun. Math. Phys.",
    volume = "333",
    number = "1",
    pages = "153--186",
    year = "2015"
}

@article{Kibble:1961ba,
    author = "Kibble, T. W. B.",
    editor = "Hsu, Jong-Ping and Fine, D.",
    title = "{Lorentz invariance and the gravitational field}",
    doi = "10.1063/1.1703702",
    journal = "J. Math. Phys.",
    volume = "2",
    pages = "212--221",
    year = "1961"
}

@article{Koivisto:2025ryb,
    author = "Koivisto, Tomi and Zheng, Lucy and Zlosnik, Tom",
    title = "{Spin(4) gauge theory of space, time, gravitation, matter, and dark matter}",
    eprint = "2507.00968",
    archivePrefix = "arXiv",
    primaryClass = "gr-qc",
    doi = "10.1103/xx3x-s5wj",
    journal = "Phys. Rev. D",
    volume = "113",
    number = "12",
    pages = "124069",
    year = "2026"
}

@book{Aldrovandi:2013wha,
    author = "Aldrovandi, Ruben and Pereira, Jos{\'e} Geraldo",
    title = "{Teleparallel Gravity}: {An Introduction}",
    doi = "10.1007/978-94-007-5143-9",
    isbn = "978-94-007-5142-2, 978-94-007-5143-9",
    publisher = "Springer",
    year = "2013"
}

@article{Zlosnik:2018qvg,
    author = "Z{\l}o{\'s}nik, Tom and Urban, Federico and Marzola, Luca and Koivisto, Tomi",
    title = "{Spacetime and dark matter from spontaneous breaking of Lorentz symmetry}",
    eprint = "1807.01100",
    archivePrefix = "arXiv",
    primaryClass = "gr-qc",
    reportNumber = "NORDITA 2018-048, NORDITA-2018-048",
    doi = "10.1088/1361-6382/aaea96",
    journal = "Class. Quant. Grav.",
    volume = "35",
    number = "23",
    pages = "235003",
    year = "2018"
}

@article{Capozziello:2026pys,
    author = "Capozziello, Salvatore and Meluccio, Giuseppe",
    title = "{The Pre-geometric Origin of Geometric Trinity of Gravity}",
    eprint = "2606.17580",
    archivePrefix = "arXiv",
    primaryClass = "gr-qc",
    month = "6",
    year = "2026"
}

@article{Stelle:1979aj,
    author = "Stelle, K. S. and West, Peter C.",
    title = "{Spontaneously Broken De Sitter Symmetry and the Gravitational Holonomy Group}",
    reportNumber = "ICTP-78-79-19",
    doi = "10.1103/PhysRevD.21.1466",
    journal = "Phys. Rev. D",
    volume = "21",
    pages = "1466",
    year = "1980"
}

@article{Gryb:2012qt,
    author = "Gryb, Sean and Mercati, Flavio",
    title = "{2+1 gravity on the conformal sphere}",
    eprint = "1209.4858",
    archivePrefix = "arXiv",
    primaryClass = "gr-qc",
    doi = "10.1103/PhysRevD.87.064006",
    journal = "Phys. Rev. D",
    volume = "87",
    number = "6",
    pages = "064006",
    year = "2013"
}

@article{Addazi:2026eto,
    author = "Addazi, Andrea",
    title = "{Topological quantization of the vacuum energy from pre-geometric gravity}",
    doi = "10.1140/epjc/s10052-026-16181-2",
    journal = "Eur. Phys. J. C",
    volume = "86",
    number = "8",
    pages = "916",
    year = "2026"
}

@article{Addazi:2026suu,
    author = "Addazi, Andrea and Meluccio, Giuseppe",
    title = "{Emergence of gravity{\textquoteright}s dynamical and topological sectors from pregeometry}",
    eprint = "2609.10619",
    archivePrefix = "arXiv",
    primaryClass = "gr-qc",
    doi = "10.1103/fytf-m83r",
    journal = "Phys. Rev. D",
    volume = "114",
    number = "4",
    pages = "044056",
    year = "2026"
}

@article{Addazi:2025vbw,
    author = "Addazi, Andrea and Capozziello, Salvatore and Marcian{\`o}, Antonino and Meluccio, Giuseppe",
    title = "{Hamiltonian analysis of pregeometric gravity}",
    eprint = "2505.01272",
    archivePrefix = "arXiv",
    primaryClass = "gr-qc",
    doi = "10.1103/v5wg-sy4w",
    journal = "Phys. Rev. D",
    volume = "112",
    number = "6",
    pages = "064058",
    year = "2025"
}

@article{Westman:2013mf,
    author = "Westman, H. F. and Zlosnik, T. G.",
    title = "{Exploring Cartan gravity with dynamical symmetry breaking}",
    eprint = "1302.1103",
    archivePrefix = "arXiv",
    primaryClass = "gr-qc",
    doi = "10.1088/0264-9381/31/9/095004",
    journal = "Class. Quant. Grav.",
    volume = "31",
    pages = "095004",
    year = "2014"
}

@article{Volovik:2023faj,
    author = "Volovik, G. E.",
    title = "{Gravity through the prism of condensed matter physics}",
    eprint = "2307.14370",
    archivePrefix = "arXiv",
    primaryClass = "cond-mat.other",
    doi = "10.31857/S1234567823190126",
    journal = "Pisma Zh. Eksp. Teor. Fiz.",
    volume = "118",
    number = "7",
    pages = "546--547",
    year = "2023"
}

@article{Sawicki:2024ryt,
    author = "Sawicki, Ignacy and Trenkler, Georg and Vikman, Alexander",
    title = "{Causality and stability from acoustic geometry}",
    eprint = "2412.21169",
    archivePrefix = "arXiv",
    primaryClass = "gr-qc",
    doi = "10.1007/JHEP10(2025)227",
    journal = "JHEP",
    volume = "10",
    pages = "227",
    year = "2025"
}

@article{Krasnov:2011pp,
    author = "Krasnov, Kirill",
    title = "{Pure Connection Action Principle for General Relativity}",
    eprint = "1103.4498",
    archivePrefix = "arXiv",
    primaryClass = "gr-qc",
    doi = "10.1103/PhysRevLett.106.251103",
    journal = "Phys. Rev. Lett.",
    volume = "106",
    pages = "251103",
    year = "2011"
}

@article{Motohashi:2016prk,
    author = "Motohashi, Hayato and Suyama, Teruaki and Takahashi, Kazufumi",
    title = "{Fundamental theorem on gauge fixing at the action level}",
    eprint = "1608.00071",
    archivePrefix = "arXiv",
    primaryClass = "gr-qc",
    reportNumber = "RESCEU-26-16",
    doi = "10.1103/PhysRevD.94.124021",
    journal = "Phys. Rev. D",
    volume = "94",
    number = "12",
    pages = "124021",
    year = "2016"
}

@article{Peldan:1993hi,
    author = "Peldan, Peter",
    title = "{Actions for gravity, with generalizations: A Review}",
    eprint = "gr-qc/9305011",
    archivePrefix = "arXiv",
    reportNumber = "GOTEBORG-93-13",
    doi = "10.1088/0264-9381/11/5/003",
    journal = "Class. Quant. Grav.",
    volume = "11",
    pages = "1087--1132",
    year = "1994"
}

@article{Bak:2022nrv,
    author = "B{\k a}k, Bart{\l}omiej and Kijowski, Jerzy",
    title = "{How the non-metricity of the connection arises naturally in the classical theory of gravity}",
    eprint = "2209.05936",
    archivePrefix = "arXiv",
    primaryClass = "gr-qc",
    doi = "10.1063/5.0208497",
    journal = "J. Math. Phys.",
    volume = "65",
    number = "9",
    pages = "092501",
    year = "2024"
}

@article{Starobinsky:1980te,
    author = "Starobinsky, Alexei A.",
    editor = "Khalatnikov, I. M. and Mineev, V. P.",
    title = "{A New Type of Isotropic Cosmological Models Without Singularity}",
    doi = "10.1016/0370-2693(80)90670-X",
    journal = "Phys. Lett. B",
    volume = "91",
    pages = "99--102",
    year = "1980"
}

@article{Westman:2014yca,
    author = "Westman, Hans F. and Zlosnik, T. G.",
    title = "{An introduction to the physics of Cartan gravity}",
    eprint = "1411.1679",
    archivePrefix = "arXiv",
    primaryClass = "gr-qc",
    doi = "10.1016/j.aop.2015.06.013",
    journal = "Annals Phys.",
    volume = "361",
    pages = "330--376",
    year = "2015"
}

@article{MacDowell:1977jt,
    author = "MacDowell, S. W. and Mansouri, F.",
    title = "{Unified Geometric Theory of Gravity and Supergravity}",
    reportNumber = "COO-3075-164",
    doi = "10.1103/PhysRevLett.38.739",
    journal = "Phys. Rev. Lett.",
    volume = "38",
    pages = "739",
    year = "1977",
    note = "[Erratum: Phys.Rev.Lett. 38, 1376 (1977)]"
}

@article{Albertus:2026fbe,
    author = "Albertus, Conrado and others",
    title = "{WISPedia -- the WISPs Encyclopedia}",
    eprint = "2602.09089",
    archivePrefix = "arXiv",
    primaryClass = "hep-ph",
    reportNumber = "IPPP/26/14, IFT-UAM/CSIC-26-9",
    month = "",
    year = "2026"
}

\end{document}